\documentclass[10pt,letterpaper,twocolumn]{article}
\usepackage[english]{babel}
\usepackage{graphicx}

\newcommand{\alt}{\mathbin{\lower 3pt\hbox
   {$\rlap{\raise 5pt\hbox{$\char'074$}}\mathchar"7218$}}}
\newcommand{\agt}{\mathbin{\lower 3pt\hbox
   {$\rlap{\raise 5pt\hbox{$\char'076$}}\mathchar"7218$}}}

\begin{document}

\setcounter{footnote}{0}
\setcounter{equation}{0}
\setcounter{figure}{0}
\setcounter{table}{0}

\title{\large\bf Multifractality and quantum diffusion\\
from self-consistent theory of localization }

\author{\small  I. M. Suslov \\
\small Kapitza Institute for Physical Problems,
\\  \small Moscow, Russia \\{} \\
\parbox{120mm}{\footnotesize \,Multifractal properties of wave
functions in a disordered system can be derived from
self-consistent theory of localization by Vollhardt and
W${\rm {\ddot o}}$lfle. A diagrammatic interpretation of
results allows to obtain all scaling relations used in
numerical experiments.  The arguments are given that
the one-loop Wegner result for a space dimension $d=2+\epsilon$
may appear to be exact, so the multifractal spectrum is
strictly parabolical.  The $\sigma$-models are shown to be
deficient at the four-loop level and the  possible reasons
of that are discussed. The extremely slow convergence to the
thermodynamic limit is demonstrated. The open question on
the relation between multifractality and a spatial dispersion
of the diffusion coefficient  $D(\omega,q)$ is resolved
in the compromise manner due to ambiguity of the $D(\omega,q)$
definition. Comparison is made with the extensive numerical
material.   } }

\date{}

\maketitle

\begin{center}
{\bf 1. Introduction}
\end{center}

In previous papers \cite{1,2,3,4,5}  we have initiated
a systematic analysis  of numerical
algorithms used in the Anderson transition studies \cite{6}.
Suggesting validity of self-consistent theory of
localization by Vollhardt and W${\rm {\ddot o}}$lfle \cite{7},
we have derived the finite-size scaling equations for the
minimal Lyapunov exponent \cite{1}, the mean conductance
\cite{2} and level statistics \cite{3}. Comparison with
numerical results shows \cite{1,2,3,4,5} that on the level of raw
data they are perfectly compatible with the self-consistent
theory, while the opposite statements of the original papers
are related with ambiguity of interpretation. It
gives a serious support to arguments \cite{100,17} that
 the Vollhardt and W${\rm {\ddot o}}$lfle
theory predicts the exact critical behavior.

The present paper deals with the next algorithm based on
the finite-size scaling for inverse participation ratios
\cite{6}, which are defined as
$$
P_q=\int d^dr |\Psi({\bf r})|^{2q} \,,
 \eqno(1)
 $$
where $\Psi({\bf r})$ is a normalized wave function of an
electron in a finite disordered system having a form of the
$d$-dimensional cube with a side $L$. In the metallic state,
the wave function  $\Psi({\bf r})$ extends along the whole
system and the normalization condition gives  $|\Psi({\bf
r})|^2\sim L^{-d}$ and $P_q\sim L^{-d(q-1)}$. In the critical
region, the wave functions acquire multifractal properties,
so
$$
\langle P_q\rangle \sim L^{-D_q(q-1)} \sim
L^{-d(q-1)+\Delta_q}\,
 \eqno(2)
 $$
and the geometrical dimension $d$ is replaced by a set
of fractal dimensions $D_q$. According to Wegner \cite{8},
the following result takes place for a space dimension
$d=2+\epsilon$
$$
\Delta_q = q(q-1) \epsilon + O(\epsilon^4 )\,,
 \eqno(3)
 $$
so the spectrum of anomalous dimensions $\Delta_q$ is
parabolic in the first  $\epsilon$-approximation.

The fractal dimensions  $D_q$ determine the behavior of
certain correlators; in particular,
$$
\langle |\Psi({\bf r})|^2 |\Psi({\bf r}')|^2\rangle \sim
|{\bf r}-{\bf r}'|^{-\eta} \,,
\eqno(4)
$$
where
$$
\eta = d-D_2 \,.
 \eqno(5)
 $$
Equation (4) is valid in the critical region  $L\alt \xi$,
where $\xi$ is the correlation length. In the metallic phase, such
behavior  persists on the scales $|{\bf r}\!-\!{\bf r}'|\alt
\xi$, while the constant limit is reached for $|{\bf r}\!-\!{\bf
r}'|\agt \xi$. In the dielectric region, dependence (4) is valid
for $|{\bf r}\!-\!{\bf r}'|\alt \xi$ and changes by exponential
decreasing for $|{\bf r}\!-\!{\bf r}' |\agt \xi$.  Since
integration of (4)  over ${\bf r}$ and ${\bf r}'$ gives unity,
one can estimate the  proportionality constant in the right hand
side and obtain for  $P_2$
$$
\langle P_2\rangle \sim
\left \{ \begin{array}{cc}
L^{-d} \left(\xi/a \right)^\eta & \mbox{(metal)} \\
L^{-d} \left(L/a \right)^\eta & \mbox{(critical region)} \\
\xi^{-d} \left(\xi/a \right)^\eta & \mbox{(dielectric)}
 \end{array} \right.\,,
 \eqno(6)
 $$
where $a$ is an atomic scale. Three results in (6)
match at $\xi\sim L$, and a comparison with (2) leads
to relation (5).

It is usually accepted \cite{6} that beyond the critical point
Eq.2 is replaced by the following relation
$$
\langle P_q\rangle = L^{-D_q(q-1)} F\left(L/\xi
\right) \,,
\eqno(7)
$$
which can be used for investigation of the critical behavior
of $\xi$. Below we show
that self-consistent theory of localization
allows to reproduce results  (2--7) and obtain all functional
relations in the explicit form. The calculated scaling functions
can be compared with the extensive numerical material.
Analogously to \cite{1,2,3,4,5}, it appears that the raw
numerical data are perfectly compatible with the Vollhardt and
W${\rm {\ddot o}}$lfle theory, while the opposite statements of
the corresponding authors are related with ambiguity of
interpretation and existence of small parameters of the Ginzburg
number type.

According to certain authors \cite{9,10}, a spatial
dispersion of the diffusion coefficient $D(\omega,q)$ is
also related with multifractal properties. The diffusion
constant  $D_L$ of a finite system of size $L$ is
determined for a given function $D(\omega,q)$ by
the relation
$$
D_L\sim D\left( D_L/L^2, L^{-1} \right) \,.
\eqno(8)
$$
If the power law dependence in  $\omega$ and  $q$ is accepted,
then it is easy to see that a combination
$$
D(\omega, q)  \sim \omega^{\eta'/d} q^{d-2-\eta'}
\eqno(9)
$$
provides the correct behavior $D_L\sim L^{2-d}$ at the critical
point \cite{11} for an arbitrary value of the exponent $\eta'$.
The hypothesis put forward by Chalker \cite{9} suggests an
equality $\eta'=\eta$, supported in \cite{9,10} by a detailed
numerical analysis. In our opinion, these arguments are
logically deficient: this fact was stressed in \cite{12}, but no
constructive alternative was suggested.

On the other hand,  attempts to introduce a spatial dispersion
into the scheme of self-consistent theory of localization
\cite{12a,101} reveal the utmost undesirability of this
modification. In absence of a spatial dispersion,
the theory possesses a lot of merits:

(a) it provides  the Wegner relation $s=\nu(d-2)$
between critical exponents of conductivity ($s$) and the
correlation length ($\nu$);

(b) it gives the correct value of the upper critical dimension
$d_{c2}=4$, which is a rigorous consequence of the
Bogoliubov theorem \cite{102} on renormalizability of  $\phi^4$
theory \cite{1,5};

(c) it gives the correct dependence $D(\omega,0)\sim
\omega^{(d-2)/d}$ at the critical point, which can be obtained by
different methods  \cite{13,14,15} and was confirmed numerically
 \cite{16};

(d) it provides a consistent description of finite
systems considered as zero-dimensional \cite{2}.

Appearance of a spatial dispersion immediately destroys
all properties  (a--d) \cite{2,5}:
it hardly can be considered as incident, since the
Vollhardt and W${\rm {\ddot o}}$lfle theory is at least a
very successful approximation. In fact, absence of an
essential spatial dispersion of  $D(\omega,q)$ was established
by the present author \cite{17} in the result of a detailed
analysis.

This contradiction can be resolved in the compromise manner,
since a definition of  $D(\omega,q)$
is ambiguous and allows the "gauge transformation" \cite{17}. A
spatial dispersion is absent in the "natural" gauge used in
\cite{17}, but it arises in other gauges
allowing the equality  $\eta'=\eta$.  Unfortunately, it makes
unclear what gauge corresponds to the observable diffusion
coefficient; there are indications that in this case the equality
$\eta'=\eta$ is violated  (Sec.6)

\begin{center}
{\bf 2. Two-point correlator}
\end{center}

\begin{center}
{\small\bf 2.1. Diagrammatic analysis}
\end{center}

Consider the correlator of two local densities of
states
$$
K_{E+\omega,E}({\bf r},{\bf r}') =
\langle \nu_{E+\omega}({\bf r})\nu_{E}({\bf r}') \rangle \, =
$$
$$
=\left\langle \sum\limits_{s,s'}\, |\psi_{s}({\bf r})|^2
|\psi_{s'}({\bf r}')|^2 \delta(E+\omega-\epsilon_s)
\delta(E-\epsilon_{s'})\right\rangle \,
\eqno(10)
$$
($\psi_s({\bf r})$ and $\epsilon_s$ are  exact eigenfunctions
and eigenenergies  for an electron in a random potential),
which is closely related with correlator (4) and can be expressed
in terms of two-particle Green functions
$$
K_{E+\omega,E}({\bf r},{\bf r}') =
\frac{1}{2\pi^2 }\,
{\rm Re} \, \left[\Phi^{RA}
({\bf r},{\bf r},{\bf r}',{\bf r}')\right.-
$$
$$
\left.
-\Phi^{RR}({\bf r},{\bf r},{\bf r}',{\bf r}')
\right] \,.
\eqno(11)
$$
Here
$$
\Phi^{RA}({\bf r}_1,{\bf r}_2,{\bf r}_3,{\bf r}_4)=
 \left\langle G^{R}_{E+\omega}({\bf r}_1,{\bf r}_2)
    G^{A}_{E}({\bf r}_3,{\bf r}_4)  \right\rangle
\eqno(12)
$$
and $\Phi^{RR}$ is defined analogously. Practically,
the diagrammatic technique is applied to the quantity
$\Phi^{RA}_{\bf kk'}({\bf q})$ (Fig.1), which is the Fourier
transform of (12) with the three-momenta designations taken
into account
\begin{figure*}
\centerline{\includegraphics[width=6.4 in]{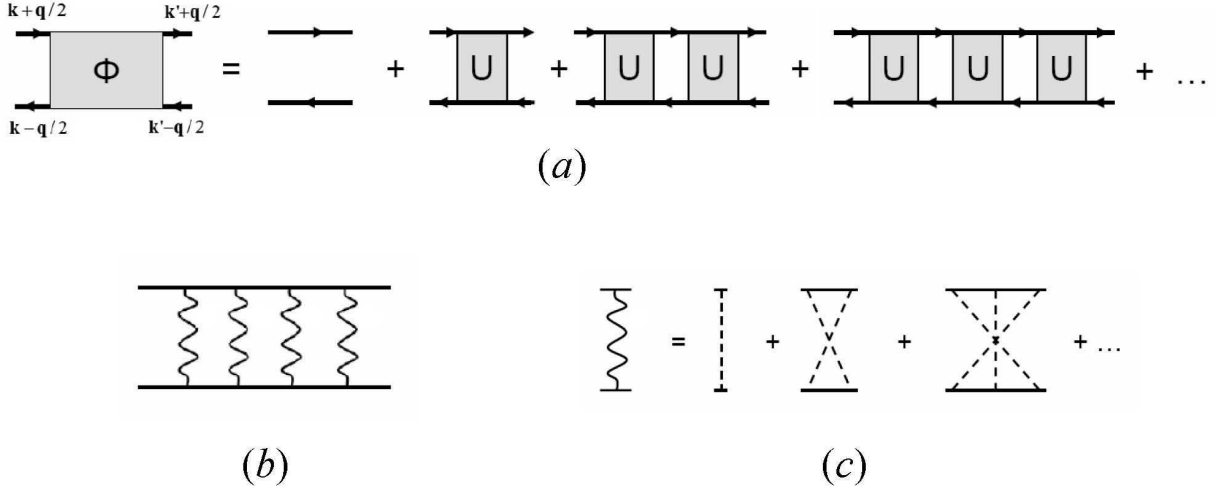}} \caption{
$(a)$ Relation of the function  $\Phi^{RA}_{\bf kk'}({\bf q})$
with the irreducible vertex $U^{RA}_{\bf kk'}({\bf q})$; $(b)$
the ladder diagrams; $(c)$ a definition of the "cooperon".  }
\label{fig1}
\end{figure*}
$$
\Phi^{RA}({\bf r}_1,{\bf r}_2,{\bf r}_3,{\bf r}_4)=
L^{-3d} \, \sum\limits_{\bf k,k',q} \,
\Phi^{RA}_{\bf kk'}({\bf q})  \cdot
$$
$$
\cdot e^{i{\bf k}\cdot ({\bf r}_1-{\bf r}_3)
+i{\bf k}'\cdot ({\bf r}_4-{\bf r}_2)+
i {\bf q}\cdot ({\bf r}_1-{\bf r}_2+{\bf r}_3-{\bf r}_4)/2}
\,.
\eqno(13)
$$
The quantity  $\Phi^{RR}$  contains no diffusion poles
and its contribution is only essential in the zero order
over the random potential. The quantity  $\Phi^{RA}$ is determined
by the irreducible four-leg vertex $U^{RA}$ (Fig.1,a), which
reduces to the "cooperon" (Fig.1,c) in the metallic phase
\cite{103}
$$
U^{C}_{{\bf k} {\bf k}^\prime} ({\bf q}) =
 \frac{2 U_0 \gamma }{-i \omega  + D_0 ({\bf k} + {\bf
k}^\prime)^2} \, \equiv U({\bf k} + {\bf k}') \,.
\eqno(14)
$$
The full $U$-vertex differs from (14) by the replacement of the
classical value  $D_0$ by the exact diffusion coefficient
$D(\omega,q)$ (see Sec.6); here
$U_0=W^2 a^d$, $W$ is an amplitude of the random potential,
$\gamma=\pi U_0 \nu_F$ is an elastic attenuation, determined
by the relation $\gamma=-{\rm Im} \Sigma^R_{\bf k}$ in terms of
the average Green function\,\footnote{\,Below we omit  signs of
averaging and accept the energy variable to be equal $E+\omega$
for functions $G^{R}$ and  $E$ for functions $G^{A}$.}
$$
 \langle G^{R}_{\bf k} \rangle   =
 \frac{1 }{E-\epsilon_{\bf k} -\Sigma^R_{\bf k}}
 \,,
$$
and $\nu_F$ is the density of states at the Fermi level. In
particular, the one-cooperon contribution to correlator (10)
has a form
$$
K_{E+\omega,E}^{(1)}({\bf r},{\bf r}') =\frac{1}{2\pi^2 }\,
{\rm Re} \,L^{-3d} \,
\sum\limits_{{\bf k},{\bf q},{\bf q}_1} P_{{\bf k}}({\bf q})
\cdot
$$
$$
\cdot \frac{2 U_0 \gamma}{-i\omega+D_0  q_1^2}\,
P_{{\bf -k+q}_1}({\bf q}) e^{i(2{\bf k}-{\bf q}_1)\cdot ({\bf
r}-{\bf r}')} \,
\eqno(15)
$$
(where  $P_{\bf k}({\bf q}) =G^{R}_{{\bf k+q}/2} G^{A}_{{\bf
k-q}/2}$) and is easily calculated in the "pole approximation",
when momenta like ${\bf q}_1$ entering the diffusion denominators
are neglected in  slowly varying functions of type
$P_{{\bf -k+q}_1}({\bf q})$. In this approximation, one can
easily calculate contributions to  $\Phi^{RA}$ from the ladder
diagrams shown in Fig.1,b, which have a qualitatively different
behavior for even ($2n$) and odd ($2n\!+\!1$) number of
cooperons\,\footnote{\,To retain symmetry of
$\Phi^{RA}({\bf r}_1,{\bf r}_2,{\bf r}_3,{\bf r}_4)$ relative
to permutation of ${\bf r}_3$ and ${\bf r}_4$, we
have added the contributions of diagrams with the
reversed lower  $G$-line.}:
$$
\Phi^{(2n)}({\bf r}_1,{\bf r}_2,{\bf r}_3,{\bf r}_4) =
 2\, k_n({\bf r}_1\!-\!{\bf r}_2) \,
k^*_n({\bf r}_3\!-\!{\bf r}_4) \cdot
$$
$$
\qquad\qquad\qquad\cdot \left[ U({\bf r}_1\!-\!{\bf
r}_3))\right]^{2n} \,, \eqno(16)
$$
$$
\Phi^{(2n+1)}({\bf r}_1,{\bf r}_2,{\bf r}_3,{\bf r}_4) = \left[
\tilde k_n({\bf r}_1\!-\!{\bf r}_4)\, \tilde k_n({\bf
r}_2\!-\!{\bf r}_3) +  \right.
$$
$$
\left.
+\tilde k_n({\bf r}_1\!-\!{\bf r}_3)
\,\tilde k_n({\bf r}_2\!-\!{\bf r}_4) \right]
\left[ U({\bf r}_1\!-\!{\bf r}_2))\right]^{2n+1} \,.
$$
Here  $U({\bf r})$ is the reverse Fourier transform of
(14),
$$
U({\bf r}) = L^{-d} \sum\limits_{{\bf q}}  \frac{2 U_0 \gamma}
{{-i \omega  + D_0 q^2}} e^{i {\bf q}\cdot {\bf r}} \,\propto \,
\frac {1}{r^{d-2}},
\eqno(17)
$$
while $k_n({\bf r})$ and $\tilde k_n({\bf r})$ are
short-ranged functions defined as
$$
\tilde k_n({\bf r}) = L^{-d} \sum\limits_{\bf k}
\left( G^{R}_{{\bf k}} G^{A}_{{\bf k}} \right)^{n+1}
e^{i {\bf k}\cdot {\bf r}} \,,
$$
$$
 k_n({\bf r}) = L^{-d} \sum\limits_{\bf k}
\left( G^{R}_{{\bf k}} \right)^{n+1}
\left(G^{A}_{{\bf k}} \right)^{n}  e^{i {\bf k}\cdot {\bf r}}
\eqno(18)
$$
and decreasing as  $\exp(-r/l)$ on the mean free path $l$,
which has the atomic scale near the Anderson transition.
The function
$\Phi^{RA}({\bf r}_1,{\bf r}_2,{\bf r}_3,{\bf r}_4)$
is exponentially small, if all  ${\bf r}_i$ are
essentially different, and the long-range tails arise only in
the case of pairwise coinciding arguments: the case ${\bf
r}_1={\bf r}_2,\,\, {\bf r}_3={\bf r}_4$ corresponds
to correlator (10), while the case  ${\bf
r}_1={\bf r}_3,\,\, {\bf r}_2={\bf r}_4$ (or equivalently
 ${\bf r}_1={\bf r}_4,\,\, {\bf r}_2={\bf r}_3$) corresponds
to another correlator
$$
{\cal K}_{E+\omega,E}({\bf r},{\bf r}') =
\left\langle
\sum\limits_{s,s'}\, \psi_s({\bf r})\, \psi_s({\bf r}') \,
\psi_{s'}({\bf r}) \,\psi_{s'}({\bf r}') \,\cdot
\right.
$$
$$
\left. \cdot
\delta(E+\omega-\epsilon_s) \delta(E-\epsilon_{s'})
\vphantom{\sum}
\right\rangle \,.
\eqno(19)
$$
According to (16), the long-range tails of correlators $
K_{E+\omega,E}({\bf r},{\bf r}')$ and ${\cal K}_{E+\omega,E}({\bf
r},{\bf r}')$ are determined by even and odd orders
correspondingly.

Using values of functions $k_n({\bf r})$ and $\tilde k_n({\bf
r})$ at zero (the energy dependence of the density of
states $\nu(\epsilon)$ is neglected)
$$
 k_n(0) = -\frac{i\nu_F}{\gamma^{2n}} a_n \,,\quad
\tilde k_n(0) = \frac{\nu_F}{\gamma^{2n+1}} a_n\,,
\eqno(20)
$$
$$
a_n = \int_{-\infty}^\infty \frac{dx\quad}{(x^2+1)^{n+1}}
= \frac{\Gamma(1/2) \Gamma(n+ 1/2)}{\Gamma(n+1)}
$$
one has for the essential contributions to (10)
$$
K^{(2n)}({\bf r},{\bf r}') = \frac{a_n^2}{\pi^2} \,\nu_F^2 \,
{\rm Re} \, \left[ \frac{2}{\pi \nu_F D(\omega)}
\Pi({\bf r}\!-\!{\bf r}'))\right]^{2n}
\eqno(21)
$$
and to (19)
$$
{\cal K}^{(2n+1)}({\bf r},{\bf r}') = \frac{a_n^2}{2\pi^2}
\,\nu_F^2 \, {\rm Re} \, \left[ \frac{2}{\pi \nu_F D(\omega)}
\Pi({\bf r}\!-\!{\bf r}'))\right]^{2n+1}
\eqno(22)
$$
where
$$
\Pi({\bf r}) = L^{-d} \sum\limits_{{\bf q}}
\frac{e^{i {\bf q}\cdot {\bf r}}}
{{ q^2 + m^2}}  \,,
$$
$$
  m^2 \equiv (-i\omega)/D(\omega) =
\xi_{0D}^{-2} \,.
\eqno(23)
$$
The replacement of $D_0$ by $D(\omega)$ extends the
formula obtained for the metallic phase to the
whole range of parameters, since it corresponds
to the replacement of the cooperon lines  (Fig.1,b,c) by the
$U$-vertices  (Fig.1,a); the ladder diagrams are sufficient, since
all diagrams has the ladder form in terms of the  $U$ blocks.
In correspondence with  \cite{17}, we neglect the $q$-dependence
of the diffusion coefficient, which is inessential in the
gauge assumed here (Sec.6). In a closed finite system, the
diffusion constant has the localization behavior
$D(\omega)= -i \omega \xi_{0D}^2$, where $\xi_{0D}$ if the
correlation length of the corresponding quasi-zero-dimensional
system \cite{2}, so the quantity $m^2$ in (23) is finite.
Transition to open systems leads to appearance of the
effective damping $\gamma_0$, which is introduced by a
change $-i \omega \longrightarrow -i \omega +\gamma_0$
simultaneously in $-i \omega$ and in $D(\omega)$; as a result,
the finite diffusion constant $\gamma_0 \xi_{0D}^2$ arises in the
static limit and the sign of the real part can be omitted in (21)
and (22).

\begin{center}
{\small\bf 2.2. Insufficiency of the pole approximation}
\end{center}

Contributions (21) can be easily summed over $n$,
$$
K({\bf r},{\bf r}')=\frac{\nu_F^2}{\pi}\,
{\rm Re}\,\int_{-\infty}^\infty
\frac{dt}{\sqrt{1+t^2} \sqrt{1+t^2-u^{2}}  } \,,
$$
$$
\qquad u=\frac{2}{\pi \nu_F D(\omega)} \Pi({\bf r}\!-\!{\bf r}')
\,,
 \eqno(24)
 $$
if $a_n^2$ is represented as the double integral (see (20)).
However, this result is practically useless due to insufficiency
of the pole approximation. In order to clarify a situation,
let estimate a value of (21) for  ${\bf r}={\bf r}'$.
Using the Ward identity \cite{7}
$$
\Delta \Sigma_{\bf k} ({\bf q}) = L^{-d} \sum \limits_{{\bf k}'}
U_{{\bf kk}'} ({\bf q}) \Delta G_{{\bf k}'} ({\bf q}) \,,
 \eqno(25)
$$
$$
\Delta G_{\bf k} ({\bf q}) \equiv G_{{\bf k} +
{\bf q}/2}^R - G_{{\bf k} - {\bf q}/2}^A,
\quad \Delta \Sigma_{\bf k} ({\bf q}) \equiv
\Sigma_{{\bf k} + {\bf q}/2}^R -
\Sigma_{{\bf k} - {\bf q}/2}^A
$$
and the relation
$$
\Delta G_{\bf k} ({\bf q}) = \left[-\omega +
\epsilon_{{\bf k+q}/2} - \epsilon_{{\bf k-q}/2}
+ \Delta \Sigma_{\bf k} ({\bf q})
\right] P_{\bf k} ({\bf q}) \,,
$$
one has for ${\bf q}=0$
$$
L^{-d} \sum \limits_{{\bf k}'}
U_{{\bf kk}'} (0) P_{{\bf k}'} (0)= 1+i\omega/2\gamma
 \eqno(26)
$$
if ${\rm Im} \Sigma^R_{\bf k} = -\gamma$ is assumed to be
independent of ${\bf k}$.\,\footnote{\,The ${\bf k}$ dependence
of $\Sigma^R_{\bf k}$  has no qualitative significance:  in
particular, it is rigorously absent in the Lloyd model, which is
quite ordinary from the viewpoint of the Anderson transition. In
the general case, neglecting of the ${\bf k}$ dependence
corresponds (in the coordinate representation) to the replacement
of short-range contributions by the $\delta$-functional ones
(see Sec.3). } For small ${\bf q}$ and the vertex $U_{{\bf kk}'}
({\bf q})$, independent of momenta, the following relation takes
place \cite{103}

$$
L^{-d} \sum \limits_{{\bf k}'}
U_0 P_{{\bf k}'} ({\bf q})= 1+i\omega \tau -D_0\tau q^2
\,,
\quad \tau=1/2\gamma .
 \eqno(27)
$$
There are serious grounds to expect the analogous relation
in the general case
$$
L^{-d} \sum \limits_{{\bf k}'}
U_{{\bf kk}'} ({\bf q}) P_{{\bf k}'} ({\bf q})=
1+i\omega \tau -D(\omega,q)\tau q^2\,.
\eqno(28)
$$
Indeed, the right hand side of (25), being a function of
${\bf k}$ and ${\bf q}$, can be practically specified
as a function of invariants  $k^2$, $q^2$, ${\bf k\cdot q}$
and allows an expansion over the second and the third of
them. In the absence of the ${\bf k}$  dependence in the
left hand side of (25), one can average over directions of
 ${\bf k}$ and remove the odd orders in ${\bf k\cdot q}$.
 As a result, the right hand side of (28) contains only
 even orders in ${\bf  q}$. A zero order term is specified by
(26), while the higher orders can be absorbed by a
definition of $D(\omega,q)$. The slow dependence on the modulus
of  ${\bf k}$ is removed by estimation at $k^2\approx \epsilon_F$.

The equality  ${\bf r}={\bf r}'$  in (11) leads to ${\bf r}_2={\bf
r}_4$ in (13), so $\Phi_{{\bf kk}'} ({\bf q})$
enters in the form summed over ${\bf k}'$ and the radical
simplifications are possible due to (28).  For example,
one has for the diagram with two  $U$ blocks
$$
L^{-d} \sum \limits_{{\bf k}'}  \Phi^{(2)}_{{\bf kk}'} ({\bf q}) =
$$
$$
=L^{-2d} \sum \limits_{{\bf k}_1 {\bf k}'} P_{{\bf k}} ({\bf q})
U_{{\bf kk}_1} ({\bf q}) P_{{\bf k}_1} ({\bf q})
U_{{\bf k}_1{\bf k}'} ({\bf q}) P_{{\bf k}'} ({\bf q})=
$$
$$
=L^{-d} \sum \limits_{{\bf k}_1} P_{{\bf k}} ({\bf q})
U_{{\bf kk}_1} ({\bf q}) P_{{\bf k}_1} ({\bf q})
 e^{ i\omega \tau -D(\omega,q)\tau q^2 } =
$$
$$
=  P_{{\bf k}} ({\bf q})
 e^{ 2 \left[ i\omega \tau -D(\omega,q)\tau q^2 \right] }  \,.
 \eqno(29)
$$
Analogously, in the case of $n$ blocks
$$
L^{-d} \sum \limits_{{\bf k}'}  \Phi^{(n)}_{{\bf kk}'} ({\bf q}) =
  P_{{\bf k}} ({\bf q})
 e^{ n \left[ i\omega \tau -D(\omega,q)\tau q^2 \right] }
 \eqno(30)
$$
and summation over $n$ gives
$$
L^{-d} \sum \limits_{{\bf k}'}  \Phi_{{\bf kk}'} ({\bf q}) =
  P_{{\bf k}} ({\bf q}) \frac{2\gamma}
{-i\omega  +D(\omega,q) q^2} =
$$
$$
=   \frac{i\Delta G_{\bf k} ({\bf q}) + O(q)}
{-i\omega  +D(\omega,q) q^2} \,.
\eqno(31)
$$
The exactly such relation follows from the Bethe-Salpeter equation
(see formula (63) in \cite{17}), so the introduced function
$D(\omega,q)$ can be identified with the diffusion coefficient.

Substitution of (30)  in (10,13) gives for the $n$-th order
contribution
$$
K^{(n)}(0,0) = \frac{1}{\pi^2}{\rm Re} \,
L^{-2d} \sum \limits_{{\bf
k},{\bf q}} P_{{\bf k}} ({\bf q})
\cdot
$$
$$
\cdot e^{ n \left[ i\omega \tau
 -D(\omega,q)\tau q^2 \right] } =\nu_F^2 \,,
 \eqno(32)
 $$
where the latter equality is valid for  $\omega\to 0$ in
the vicinity of the critical point, since  $D(\omega,q)$ turns to
zero simultaneously for all $q$ \cite{17}. Introducing
dimensionless conductance $g=\nu_F D L^{d-2}$ \cite{11} and using
the relation  $g\sim (L/\xi)^{d-2}$ valid in the metallic
phase \cite{14}, one has from  (21)
 $$
 K^{(2n)}({\bf r},0) \sim \nu_F^2 \left( \xi/r
 \right)^{2n(d-2)}  \,.
 \eqno(33)
 $$
The  contributions for different $n$ become comparable for $r \sim
\xi$ and have the order of $\nu_F^2$ in correspondence with  (32).
It is clear that result (33) is valid for $r\agt \xi$, while
the $r$ dependence is saturated for  $r\alt \xi$. The latter
is a consequence of the delicate cancellations in the Ward identity
(25):  summation over ${\bf k}'$ in the infinite limits removes
the pole part of $U_{{\bf kk}'} ({\bf q})$ due to its
orthogonality to the function $\Delta G_{\bf k'} ({\bf q})$
\cite{17}; so the pole approximation is completely inapplicable.
For large $|{\bf r}- {\bf r}'|$, summation over ${\bf k}'$ is
effectively restricted by the range $|{\bf k}'|\alt |{\bf r}-{\bf
r}'|^{-1}$ and the orthogonality becomes inessential,
restoring validity of the pole approximation.

The whole correlator $ K({\bf r},0)$ is determined by the
two cooperon contribution for  $r\agt \xi$, while summation of
the series of approximately equal terms is necessary for $r\alt
\xi$:  the expansion
parameter $u$ tends to unity for  $|{\bf r}-{\bf r}'|\to 0$
and the integral (24) diverges due to the logarithmic
singularity at $u=1$. The specific form of divergency
is determined by the character of saturation of $ K^{(2n)}({\bf
r},0)$; for example, the power law behavior
$ K({\bf r},0) \sim r^{-2/\gamma}$ arises, if
$$
K^{(2n)}({\bf r},0) = \nu_F^2
 \left( 1-\beta_n r^2/\xi^2 \right)\,, \qquad     r\alt \xi
 \eqno(34)
 $$
with coefficients $\beta_n\sim n^\gamma$.
In fact, the dependence  $ K({\bf r},0) \sim r^{-\eta}$
is expected from the analogy of (4) and  (10) (Sec.2.4),
so
$$
K({\bf r},0) - \nu_F^2\,
\sim \left \{ \begin{array}{cc}
\nu_F^2 \left(\xi/r \right)^\eta, & r\alt \xi \\
{\qquad} & {\qquad}  \\
\nu_F^2 \left(\xi/r \right)^{2(d-2)}, & r\agt \xi
\end{array} \right.\,.
\eqno(35)
$$
where the zero-order contribution is separated.

In case of the correlator  $ {\cal K}({\bf r},{\bf r}')$,
the equality  ${\bf r}_2={\bf r}_4$ is valid in (13) for arbitrary
${\bf r}$ and ${\bf r}'$, and the pole approximation is
deficient from the very beginning; one always
has the result (32), while (22) has no range of
applicability\,\footnote{\,By this reason, the
argumentation of Sec.2.3 referring to the correlator
${K}({\bf r},{\bf r}')$  cannot be applied to
${\cal K}({\bf r},{\bf r}')$. }.  The use  of (31) leads to the
functional form
$$
{\cal K}({\bf r},{\bf r}') = \frac{1}{2\pi^2} \,{\rm Re} \,
L^{-d} \sum \limits_{{\bf q}}  \frac{ 2\pi\nu_F}
{-i\omega  +D(\omega,q) q^2} \, e^{i {\bf q}\cdot ({\bf r}-{\bf
r}')}  \,,
\eqno(36)
$$
corresponding to the first order contribution  (see (22)), but
the diffusion coefficient is defined in the different gauge
allowing a spatial dispersion (Sec.6).


\begin{center}
{\small\bf 2.3. Situation for $d=2+\epsilon$.}
\end{center}

In the spatial dimension $d=2+\epsilon$ with $\epsilon\ll 1$ one
has for the expansion parameter in (21,22)
$$
u = \frac{2 \Pi(r)}{\pi \nu_F D}
\sim \frac{1}{ \nu_F D} \int
\frac{d^dq}{q^2} e^{i {\bf q}\cdot {\bf r}} \sim
$$
$$
\sim
\frac{L^\epsilon}{g} \int\limits_{1/L}^{1/r} q^{-1+\epsilon}  dq
\sim \frac{1}{g} \frac{(L/r)^\epsilon-1}{\epsilon}
\eqno(37)
$$
where $g=\nu_F D L^{d-2}$ is a dimensionless conductance and
the condition  $m\alt L^{-1}$ is accepted, which is valid in the
metallic state and the critical region. We have in mind that a
perturbation theory is constructed for open systems, where
${\bf q}=0$ is not allowed value for the momentum  ${\bf q}$
and the diffusion constant is finite in the static limit
\cite{2}. Accepting $r$ satisfying the condition $\epsilon
\ln(L/r) \ll 1$ and having in mind that a value of $g$ at
the Anderson transition is $g_c\sim 1/\epsilon$, one can see
that the expansion parameter $u=\ln(L/r)/g$ is small for
the interval $L\exp(-1/\epsilon) \alt r \le  L$ both in the
metallic and critical region.  The limiting value (29) is not
attained and the two-cooperon expression is valid for correlator
(35). This expression is not affected by variation  of the
correlation length  $\xi$, which runs in the metallic phase
from the minimal value $\xi_{min}$ till infinity, and hence
$\xi$ is not manifested as a significant length scale.
Then one can conclude from  (35) that
$$
\eta =2\epsilon
  \eqno(38)
  $$
in correspondence with the Wegner result (see  (2--5)).
We do not expect that a character of the solution changes
at a scale different from $\xi$, so the restriction
$\epsilon \ln(L/r) \ll 1$ is not essential and the
two-cooperon behavior persists in the metallic phase
for arbitrary  $r$:
$$
K({\bf r},{\bf r}') -\nu_F^2 =  \,\nu_F^2 \, {\rm Re} \,
\left[ \frac{L^\epsilon }{\pi g}
\Pi({\bf r}\!-\!{\bf r}')\right]^{2}
\,.
\eqno(39)
$$
In the localized phase one has $m=\xi^{-1}$ and the expansion
parameter is $u=\ln(\xi/r)/g$, so the two-cooperon behavior
holds for  $\xi\exp(-g) \alt r \le L$. On the other hand,
for $ r \alt \xi$ we expect the same power law,
as in the critical region. Therefore, the result (39)
can be extended to the localized phase.  \vspace{2mm}

\begin{center}
{\small\bf 2.4. Relation of correlators (10) and (4).}
\end{center}

In development of the perturbation theory the system is
assumed to be open, so its conductance   $g$ is finite
and an expansion over $1/g$ is possible. Interpretation of
expressions like (10) in open systems suggests broadening
of the $\delta$-functions to a width  $\Gamma\gg \Delta$,
where $\Delta=1/\nu_F L^d$ is a mean level spacing. Then each
sum over $s$ and $s'$ contains $\Gamma/\Delta$ terms,
and each $\delta$-function gives a factor $1/\Gamma$.
Suggesting that all terms with  $s=s'$  (and correspondingly
 $s\ne s'$) have the same statistical properties, one has for
$\omega=0$
$$
K_{E,E}({\bf r},{\bf r}') \approx \frac{1}{\Gamma \Delta}
\left\langle  |\psi_E({\bf r})|^2 |\psi_E({\bf r}')|^2
\right\rangle +
$$
$$
+ \frac{1}{\Delta^2} \left\langle  |\psi_E({\bf
r})|^2 |\psi_{E'}({\bf r}')|^2 \right\rangle \,.
\eqno(40)
$$
Assuming for estimate  that $\psi_E({\bf r})=\Psi({\bf r-R})$
with the permanent envelope $\Psi({\bf r})$ and a random
origin ${\bf R}$, one can replace averaging over disorder
by averaging
over ${\bf R}$ and obtain for the second term in (40)
$$
 \frac{1}{\Delta^2} L^{-2d} \int d^dR\, d^dR' \,
  |\Psi({\bf r-R})|^2 \, |\Psi({\bf r}'-{\bf R}')|^2
=\nu_F^2 \,,
\eqno(41)
$$
while the first term can be estimated as
$$
\nu_F^2 (\Delta/\Gamma)
\left(L/|{\bf r}-{\bf r}'| \right)^{4\alpha-d} \,,
\eqno(42)
$$
if  $\Psi({\bf r}) \sim |{\bf r}|^{-\alpha}$ and
$d/2<2\alpha<d$. In the first approximation, the
zero order contribution  $\nu_F^2$ arises from the terms with
$s\ne s'$, while the power law behavior corresponding to (4)
is determined by the terms with $s=s'$. In fact, such
decomposition is not rigorous because variations of  ${\bf R}$
and ${\bf R}'$ are not independent, so the second term in  (40)
contains a dependence on ${\bf r}-{\bf r}'$ (see Eq.44 below).

A situation is more transparent in the limit of closed
systems, when  $\Gamma/\Delta \to 0$. Then, for $\omega=0$, the
vicinity of energy $E$ contains (with probability $\Gamma/\Delta$)
one level with a certain number $s_0$, so only the
contribution with $s=s'=s_0$ remains in  sum (10) and the
second term in (40) vanishes. For transition from open to
closed systems one should omit the zero-order contribution
$\nu_F^2$, and then correlator (10) can be identified with
(4) apart from the constant factor. Comparison with (39)
gives
$$
A \left\langle  |\psi_E({\bf r})|^2 |\psi_E({\bf r}')|^2
\right\rangle = L^{-2d} \,
{\rm Re} \, \left[ L^\epsilon
\Pi({\bf r}\!-\!{\bf r}')\right]^{2} \,,
\eqno(43)
$$
where  $A$ is determined by the normalization
condition (Sec.4). Using the properties of the diffusion
propagator $\Pi({\bf r})$ (Sec.5), one can easily show
that result (43) corresponds to the physical
expectations on correlator (4): the power law behavior
$|{\bf r}\!-\!{\bf r}'|^{-\eta}$ taking  place for
$|{\bf r}\!-\!{\bf r}'|\alt \xi$ changes for $|{\bf r}\!-\!{\bf
r}'|\agt \xi$ by saturation  in the metallic
phase and by exponential decreasing in the localized state.
The constant limit in the metallic phase is determined by
the contribution of the term with ${\bf q}=0$, which is always
present in closed systems \cite{2}. It disappears in open
systems, leading to insignificance of the scale $\xi$ in the
metallic phase, which was discussed above. The terms with
$s=s'$ are the same in correlators (10) and (19),
providing
a ground for a hypothesis \cite{18} on the identical
behavior of these correlators in the critical
region\,\footnote{\,In the general case, their behavior is surely
different, as clear from the estimate of type  (41).}.

Consider the case of finite frequencies,  $\omega\gg \Delta$.
For $\Gamma\sim \Delta$, the first term is absent in  the
expression of type (40)  and comparison with  (39) gives
$$
L^{2d} \left\langle  |\psi_E({\bf r})|^2
|\psi_{E+\omega}({\bf r}')|^2 \right\rangle \approx
1+ {\rm Re} \, \left[ \frac{L^\epsilon }{\pi g}
\Pi({\bf r}\!-\!{\bf r}')\right]^{2} \,.
\eqno(44)
$$
The power law behavior of the propagator $\Pi({\bf r})$
persists at scales less than $L_\omega$, where
$$
L_\omega=\left(\nu_F \omega\right)^{-1/d}  \,,
\eqno(45)
$$
while for  $r\agt L_\omega$ it changes by exponential decreasing
(Sec.4);
the wave functions corresponding to energies $E$
and $E+\omega$ become statistically independent for $r\agt
L_\omega$ \cite{18}.

\vspace{2mm}

\begin{center}
{\bf 3. Many-point correlators}
\end{center}

Analogously, one can define the $n$-point correlators
$$
K({\bf r}_1,{\bf r}_2,\ldots,{\bf r}_n) =
\langle \nu_{E_1}({\bf r}_1)\, \nu_{E_2}({\bf r}_2)\,\ldots
\,\nu_{E_n}({\bf r}_n)  \rangle \,
 \eqno(46)
$$
and relate them with many-particle Green functions,
e.g. for  $n=3$
$$
K({\bf r}_1,{\bf r}_2,{\bf r}_3) = -\frac{1}{4\pi^3 }\,  {\rm Im}
\, \left[\Phi^{RAR}({\bf r}_1,{\bf r}_1, {\bf r}_2,{\bf r}_2,{\bf
r}_3,{\bf r}_3)+ \right.
$$
$$
\left.+
({\bf r}_1 \leftrightarrow {\bf r}_2)+
({\bf r}_2 \leftrightarrow {\bf r}_3)
\vphantom{A^S} \right] \,,
\eqno(47)
$$
and the correlator is determined by the diagrams with three
$G$-lines (Fig.2). Selection of diagrams is conveniently
made in the coordinate representation, where the cooperon
vertex (14) has a form
\begin{figure*}
\centerline{\includegraphics[width=6.0 in]{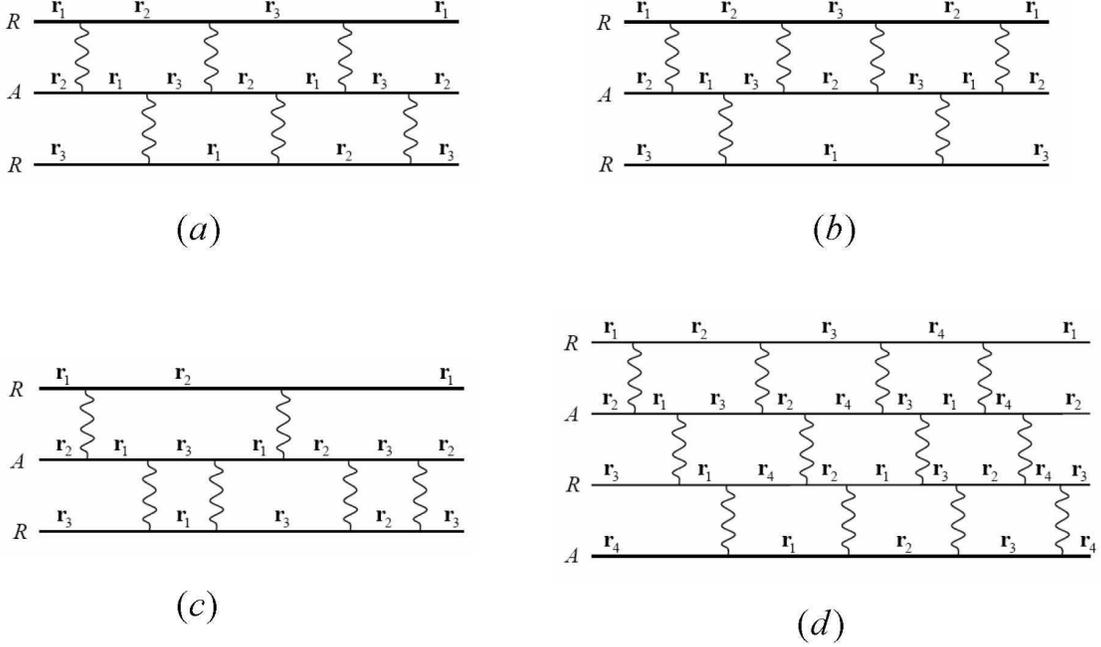}}
\caption{In the $n=3$ case, the contributions
symmetrical over all $r_{ij}$ are determined by diagrams
$(a)$, $(b)$, $(c)$ and the diagrams obtained from them by
reflections respective the horizontal and vertical axes.
$(d)$ The diagrams analogous to $(a)$
exist for arbitrary $n$.  }
\label{fig2}
\end{figure*}
$$
U^{C}({\bf r}_1,{\bf r}_2,{\bf r}_3,{\bf r}_4) =
U({\bf r}_1\!-\!{\bf r}_3) \, \delta({\bf r}_1\!-\!{\bf r}_4) \,
\delta({\bf r}_2\!-\!{\bf r}_3) \,
\eqno(48)
$$
and differs from the full four-leg vertex by the replacement
of short-range functions like  $k_n({\bf r})$ by the
$\delta$-functions. Analogously, in the analysis of power law
tails one can use the $\delta$-functions instead of the
short-range functions $G^R({\bf r})$ and $G^A({\bf r})$.
According to (48), the coordinates  ${\bf r}_i$ and ${\bf
r}_j$ corresponding to $G$-lines, coming up from the left to
a cooperon vertex, trade places after passing it
(Fig.2), and the cooperon line gives a factor $U({\bf r}_i-{\bf
r}_j)$.  Long-range contributions to correlator  (46) are
determined by diagrams, for which the coordinates of all
$G$-lines return to the same sequence after passing all
cooperon vertices.  The functions  $U({\bf r}_i-{\bf r}_j)$ can
enter only in even powers, since the coordinates ${\bf r}_i$ and
${\bf r}_j$ should be transposed even number of times to restore
the initial sequence.

Analogously to Sec.2.4, correlator (46) is related to the
$n$-point correlator of wave functions. Accepting the power law
dependence on differences ${\bf r}_{ij}={\bf r}_i-{\bf r}_j$,
one can write the most general form of such dependence
$$
\langle |\Psi({\bf r}_1)|^2 \, |\Psi({\bf r}_2)|^2\,\ldots
\,|\Psi({\bf r}_n)|^2
\rangle \sim
L^{-nd} \left(\frac{L}{a}\right)^{\kappa_n} \cdot
$$
$$
\cdot
\left[ \left(\frac{L}{r_{12}}\right)^{\alpha_n}
       \left(\frac{L}{r_{13}}\right)^{\beta_n}
       \left(\frac{L}{r_{23}}\right)^{\gamma_n} \ldots
       \left(\frac{L}{r_{n-1,n}}\right)^{\delta_n} + \right.
$$
$$
\left.   +   \mbox{\rm permutations of $r_{ij}$ }
\vphantom{A^S} \right]
      \,, \eqno(49)
$$
where permutations provide a symmetry of the expression over
all $r_{ij}$. Without loss of generality, one can accept
$$
\alpha_n \ge \beta_n \ge \gamma_n \ge \ldots
\ge \delta_n  \,.
 \eqno(50)
$$
Using (49) and the "algebra of multifractality" formulated in
\cite{18}, one can derive inequalities for  $\Delta_n$. If all
$r_{ij}\sim L$, then correlator (46) allows decomposition
\cite{18}\,\footnote{\,Since it is not quite evident,
we  accept the following procedure. Let introduce the scale
$L_\omega$ related with a frequency and defined in  (45); then
the functions $\Pi(r_{ij})$ are exponentially small for
$r_{ij}\agt L_\omega$ (Sec.4) and only the diagrams without
cooperon lines survive among diagrams like those in Fig.2; these
diagrams correspond to decomposition  (51). In case  $L_\omega
\ll L$, the latter decomposition is valid for $r_{ij}\sim L$ and
remains valid approximately, if the scale $L_\omega$ is increased
till $L$.  }
$$
\langle |\Psi({\bf r}_1)|^2\rangle \, \langle|\Psi({\bf r}_2)|^2\rangle
\,\ldots \,\langle|\Psi({\bf r}_n)|^2 \rangle \sim
L^{-nd} \,,
 \eqno(51)
$$
so $\kappa_n=0$. If all  $r_{ij}=0$, then divergencies in (49)
are cut-off on the scale  $a$, and comparing with relation
$$
\langle |\Psi({\bf r})|^{2n} \rangle \sim  L^{-nd+\Delta_n} \,
 \eqno(52)
$$
following from (1,2), one has
$$
\Delta_n=\alpha_n+ \beta_n + \gamma_n + \ldots
+ \delta_n  \,.
 \eqno(53)
$$
Using inequality (50) and taking into account that there are
$n(n-1)/2$ terms in the sum, one has $\Delta_n
\le \alpha_n\, n(n-1)/2$ which reduces to
$$
\Delta_n \le \Delta_2\, \frac{n(n-1)}{2} \,,
\eqno(54)
$$
since the exponent $\alpha_n$ is $n$-independent and
coincides with $\eta=\Delta_2$. Indeed, if  $r_{12} \ll L$ and
the rest $r_{ij} \sim L$, then (49) gives\,\footnote{\,The right
hand side of (49) may contain less singular terms, determined by
exponents  $\tilde\alpha_n$, $\tilde \beta_n$, $\ldots$, $\tilde
\delta_n$ whose sum is less than  $\Delta_n$. If it occurs that
$\tilde\alpha_n>\alpha_n$, then $\tilde\alpha_n=\eta$ and
$\alpha_n<\eta$. Hence, the inequality  $\alpha_n\le \eta$ holds
in the general case, which is sufficient for validity of (54).}
$$
\langle |\Psi({\bf r}_1)|^2 \, |\Psi({\bf r}_2)|^2\rangle
\,\langle|\Psi({\bf r}_3)|^2 \rangle
\,\ldots \,\langle|\Psi({\bf r}_n)|^2 \rangle \sim
$$
$$
\sim
L^{-nd} \left( L/r_{12}\right)^{\alpha_n} \,,
 \eqno(55)
$$
while it is proportional to $r_{12}^{-\eta}$ according to (4).
In the main $\epsilon$ approximation (see (3)),
the sign of equality takes place in
(54), and correlator (49) is determined by the most symmetric
configuration
$$
\langle |\Psi({\bf r}_1)|^2 \, |\Psi({\bf r}_2)|^2\,\ldots
\,|\Psi({\bf r}_n)|^2 \rangle \sim L^{-nd}
\prod\limits_{i<j} \left(\frac{L}{r_{ij}}\right)^{\eta}\,.
 \eqno(56)
$$
The  contribution $O(\epsilon^4)$ in Eq.3, corresponding to
the orthogonal ensemble, has a structure
$-an(n-1)(n^2-n+1$ with $a>0$ \cite{8,19}, and the
following inequality follows from (54) for $n>1$
$$
n^2-n+1 \ge 3 \,,
\eqno(57)
$$
which is satisfied for $n=2,\,3,\,4,\,\ldots$.

The present paper deals with the usual disordered systems like
electrons in a random potential, which correspond to the Dyson
orthogonal ensemble. However, inequality  (54) is not related
with self-consistent theory and has a general character. It
reduces to an equality for the parabolic spectrum $\Delta_q=\beta
q(q-1)$ with arbitrary $\beta$, leading to the symmetric form
(56) with $\eta=2\beta$ for the $n$-point correlator. In
particular, it is actual in the first $\epsilon$-approximation
for the unitary ensemble, where \cite{8,19}
$$
\Delta_n=n(n-1)(\epsilon/2)^{1/2} +
{(3/8)}  n^2(n-1)^2 \zeta(3) \epsilon^2 \,,
\eqno(58)
$$
and in the regime of the quantum Hall effect (see below).
Substitution of (58) into (54) gives the inequality for $n>1$
$$
n(n-1) \le 2   \,,
\eqno(59)
$$
which is violated for  $n=3,\,4,\,\ldots$ Hence, the results
obtained in the $\sigma$-models reveal deficiency on the
four-loop level. A possible reason of that is discussed in
Sec.7.

Expression (56) allows a diagrammatic interpretation.
For small $\epsilon$ one can neglect non-symmetrical
terms, though a mechanism of their compensation is not
quite clear.  In case $n=3$, the lowest order symmetrical
contribution arises from the diagrams in Fig.2,a--c:
$$
K({\bf r}_1,{\bf r}_2,{\bf r}_3) = {\rm const}\, \nu_F^3
\left( \frac{L^{d-2}}{g} \right)^6  \, \cdot
$$
$$ \cdot
\Pi(r_{12})^2 \, \Pi(r_{13})^2 \,\Pi(r_{23})^2 \,.
 \eqno(60)
$$
The diagrams analogous to that of Fig.2,a exist for arbitrary $n$,
as illustrated in Fig.2,d for $n=4$: the first $(n-1)$ cooperons
provide a cyclic permutation of  ${\bf r}_1,{\bf r}_2,\ldots,{\bf
r}_n$, which should be repeated  $n$ times, in order to restore
the initial configuration. Adding the zero-order term,
one has
$$
K({\bf r}_1,{\bf r}_2,\ldots,{\bf r}_n) -\nu_F^n =
\eqno(61)
$$
$$
= {\rm const}\, \nu_F^n
\left( \frac{L^{d-2}}{g} \right)^{n(n-1)}  \,
\prod\limits_{i<j} \Pi(r_{ij})^2 \,.
$$
In the limit of closed systems the term  $\nu_F^n$
disappears (Sec.2.4), and Eq.61 gives the main
symmetrical contribution in the metallic region for
$r_{ij}\agt \xi$, which can be extended to arbitrary
$r_{ij}$ analogously to Sec.2.3.

Formally, expression (61) is obtained for $d=2+\epsilon$
with small $\epsilon$, but in fact its validity is related
with two qualitative moments:

(i) insignificance of the correlation length  $\xi$ in the
metallic phase as a characteristic length scale;

(ii) realization of the maximally symmetric form  (56)
for the $n$-point correlator.

\noindent These properties can be approximate and valid only
for small  $\epsilon$.  However, their qualitative
character allows to assume that they persist in the general
case, as supported by a diagrammatic interpretation of
results.  In such a case, the multifractal spectrum is determined
by  the relation $\Delta_n=n(n-1)\epsilon$ and appears to be
strictly parabolical. Below we use Eq.61 in the $3D$ case,
considering it as extrapolation from small $\epsilon$ to
$\epsilon \sim 1$, but having in mind that it can be exact.

The simplest arguments do not allow to reject this hypothesis. A
reference to the $\epsilon$-expansion is unfounded, since
$\sigma$-models are deficient on the four-loop level (Sec.7).
Numerical data are not reliable due to  extremely slow
convergence to the thermodynamic limit (Sec.5). On the other
hand, the following arguments can be given in favor of the
hypothesis.

(a) The result $\eta=2\epsilon$ looks plausible, since the
condition $\eta>d$ is fulfilled for  $d>4$; then it follows
from (56) that the normalization integral is determined
by the atomic scale for all actual correlators. It agrees
with estimates by the optimal fluctuation method and
instanton calculations  \cite{105}, which predict localization of
wave functions for $d>4$ at the atomic scale even in
the critical region.

(b) A surprising accuracy of the Wegner one-loop result (3)
in application to the $d=3$ and $d=4$ cases was reported
in a lot of numerical experiments \cite{33,21,30,32}.
For example, a position of the maximum for the singular
spectrum  $f(\alpha)$ (which is $\alpha_0=d+\epsilon$ in the
one-loop approximation) was estimated as
$\alpha_0=4.03\pm 0.05$ \cite{33}, $\alpha_0=4.048\pm 0.003$
\cite{21} for $d=3$  and $\alpha_0=6.5\pm 0.2$ \cite{33}  for
$d=4$. A parabolic form of the spectrum is confirmed
on the level of $10\%$ \cite{21,30,32}, which should be
considered as satisfactory (Sec.5).

(c) In the regime of the integer quantum Hall effect, the spectrum
is parabolic on the level of $10^{-3}$ \cite{26}, and there are
theoretical arguments in favor of strict parabolicity
\cite{107,108,109} (confirming the property (ii)) based on
the relation with the conformal field theory.

(d) Validity of the Vollhardt and W${\rm {\ddot o}}$lfle
theory is directly related with the property (i).
Indeed, it is known from finite size scaling that
$g=g_c+const\, (L/\xi)^{1/\nu}$ in the critical region
\cite{2,6}, while $g\sim (L/\xi)^{d-2}$ in the metallic phase
\cite{14}\,\footnote{\,According to one-parameter scaling
theory  \cite{11}, $g=F(L/\xi)$ where the function $F(x)$ has a
behavior  $x^{d-2}$ in the metallic phase, in order to provide
the relation  $g\propto L^{d-2}$. Due to the dependence $\xi\sim
\tau^{-\nu}$ ($\tau$ is a distance to the transition), one can
consider  $g$ as a function of the argument  $\tau L^{1/\nu}$,
which allows the regular expansion in  $\tau$ due to the
absence of phase transitions in finite systems; the first
order in $\tau$ is sufficient in the critical region.}; it gives
the relation $\nu^{-1}=d-2$ \cite{7}, if $\xi$ is not a
significant length scale. Thereby, in the framework of
self-consistent theory the property (i) is naturally considered
as exact.

(e) Application of the "algebra of multifractality" to
correlators of the more general form than (49) leads to the
statement on  strict parabolicity of the multifractal spectrum
\cite{150}. Therefore, the symmetric form (56) is exact,
while deficiency of $\sigma$-models takes place not only
for unitary, but also for the  orthogonal ensemble.

\begin{center}
{\bf 4. Scaling for inverse participation ratios} \end{center}

We have established that the critical behavior of correlator
(56) is reproduced by the diagrammatic contribution (61). The
latter has a more wide range of applicability and allows
to extend the results beyond the critical region. In the
limit of closed systems, one has from (61) analogously to
Sec.2.4
$$
\langle |\Psi({\bf r}_1)|^2 \, |\Psi({\bf r}_2)|^2
\,\ldots \,|\Psi({\bf r}_n)|^2 \rangle =
\eqno(62)
$$
$$
 = A^{-1}  L^{-dn} \, L^{n(n-1)\epsilon}   \,
\prod\limits_{i<j} \Pi(r_{ij})^2 \,,
$$
where the parameter  $A$ is determined by the normalization
condition, since integration of the left hand side over  ${\bf
r}_1,\,\ldots,\, {\bf r}_n$ gives unity:
$$
A  = \, L^{-dn+n(n-1)\epsilon}   \, \int
d^dr_1\,\ldots\, \int d^dr_n
\prod\limits_{i<j} \Pi({\bf r}_i-{\bf r}_j)^2 \,.
\eqno(63)
$$
Integration is easily performed in case $n=2$, giving $A$
as a regular function of  $z=L/\xi_{0D}$,
$$
A=A(z)= \sum\limits_{\bf s}
\left[ \frac{1}{z^2 + (2\pi  {\bf s})^2}\right]^2\,=
$$
$$=
\left \{ \begin{array}{cc} 1/z^4, & z\ll 1 \\
\tilde c_d z^{d-4}\,, & z\gg 1
\end{array} \right.\,,
 \eqno(64)
 $$
where  $\tilde c_d=\pi K_d(1-d/2)/2\sin(\pi d/2)$,
$K_d=\left[2^{d-1}\pi^{d/2}\Gamma(d/2) \right]^{-1}$,
 and ${\bf
s}=(s_1,\ldots, s_d)$ is a vector with  integer components
 $s_i=0,\pm 1,\pm 2\ldots$. According to \cite{2}, the
 quantity $z$ is a function of the ratio  $L/\xi$
determined by the equation
$$
\pm c_d \left(L/\xi\right)^{d-2} =
 H \left(z \right) \,,
 \eqno(65)
$$
where $c_d=\pi K_d/|2\sin(\pi d /2)|$ and $H(z)$ is a function
introduced in \cite{2} with the asymptotics $1/z^2$ for  $z\ll 1$
and $-c_d z^{d-2}$  for  $z\gg 1$. Setting ${\bf r}={\bf r}'$ in
(43) and substituting to  (1), one has for $P_2$
$$
\langle P_2 \rangle = A^{-1} \,  L^{-d}   \,
\left( L/a \right)^{2\epsilon} \,,
\eqno(66)
$$
in accordance with (7). Using (64),(65), we have
$$
\langle P_2\rangle \sim
\left \{ \begin{array}{cc}
L^{-d} \left(\xi/a \right)^{2\epsilon} & \mbox{(metal)} \\
L^{-d} \left(L/a \right)^{2\epsilon} & \mbox{(critical region)}
\\ \xi^{-d} \left(\xi/a \right)^{2\epsilon} &
\mbox{(dielectric)} \end{array} \right.\,
\eqno(67)
$$
in agreement with (6).

In case of arbitrary  $n$, one can obtain from (63)
that $A(z)\sim z^{-2n(n-1)}$ in the metallic region and
$A(z)\sim z^{-d(n-1)+n(n-1)\epsilon}$ in the localized one.
The first result is a consequence of the fact that
propagator (23) in the region $m\ll L^{-1}$ is dominated
by the term with ${\bf q}=0$ and practically constant.
To obtain the second result, one changes from
variables ${\bf r}_i$ to variables ${\bf r}_1$ and ${\bf
r}'_i={\bf r}_i-{\bf r}_1$ ($i\ge 2$) and exploits the
${\bf r}_1$-independence of the integrand  and its localization
in the  $|{\bf r}'_i|\alt \xi_{0D}$ region. Then using (65) one
has
$$
A = \left \{ \begin{array}{cc}
\sim (L/\xi)^{n(n-1)\epsilon}  & \mbox{(metal)} \\
A_c\pm B (L/\xi)^{d-2}  &  \mbox{(critical region)} \\
\sim (L/\xi)^{-d(n-1)+n(n-1)\epsilon}   & \mbox{(dielectric)}
\end{array} \right.\, .
\eqno(68)
$$
Setting ${\bf r}_{ij}=0$  in (62) one gets in analogy with (66)
$$
\langle P_n \rangle \sim  A^{-1}\, L^{-d(n-1)}
\left( {L}/{a} \right)^{n(n-1)\epsilon}  \,,
\eqno(69)
$$
and
$$
\ln\langle P_n \rangle = -D_n(n-1) \ln(L/a) + {\rm const}
+F_n(L/\xi)\, \,, \eqno(70)
$$
where the  constant is chosen from
the condition $F_n(0)=0$ and
$$
F_n(x) = -\ln (A/A_c)  =
$$
$$
= \left \{
\begin{array}{cc}
-n(n-1)\epsilon \ln{x}  & \mbox{(metal)} \\
\pm B_n x^{d-2}   & \mbox{(critical region)}\\
D_n (n-1) \ln{x}     & \mbox{(dielectric)}
\end{array}
\right.\,.
\eqno(71)
$$
Evaluation of $F_n(x)$ for arbitrary  $x$ can be made
rewriting (63) in a form of the multiple sum over momenta.
Unfortunately, such expression needs tedious numerical
calculations for large $n$ and does not provide analytic
continuation to non-integer $n$. To avoid these problems,
we note that the result $A\sim
z^{-2n(n-1)}$ for the metallic phase remains valid in the
critical region by the order of magnitude; therefore, in these
regions $F_n(x)$ differs from $F_2(x)$ by a factor $n(n-1)/2$,
while in the deep of the localized phase $F_n(x)= (n-1)(D_n/D_2)
F_2(x)$.  The simplest interpolation form ensuring such
properties is as follows
$$
F_n(x) =  \left \{  \begin{array}{cc}
C_{+} F_2(\alpha x)  & \mbox{(upper branch)} \\
C_{-} F_2( x)       &  \mbox{(lower branch)}
\end{array}
\right.\,,
\eqno(72)
$$
i.e. two branches of $F_n(x)$ have the same behavior as two
branches of $F_2(x)$ and differ only by a scale transformation.
The coefficients $C_{+}$ and $C_{-}$ provide the correct
asymptotic behavior  (71) for large $x$, and the parameter
$\alpha$ ensures
symmetry of two branches for small  $x$:
$$
C_{+} = \frac{D_n(n-1)}{D_2} ,  \quad C_{-} = \frac{n(n-1)}{2}
, \quad \alpha= \left( \frac{C_{-}}{C_{+}} \right)^{1/\epsilon}
\,.
\eqno(73)
$$
\begin{figure}
\centerline{\includegraphics[width=3.0 in]{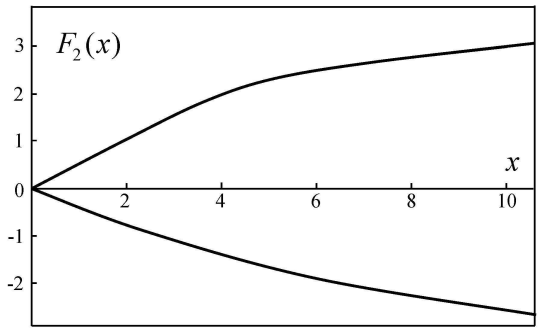}}
\caption{Scaling function  $F_n(L/\xi)$ for
$n=2$.  }
\label{fig3}
\end{figure}
If the function  $F_2(x)$ is calculated (Fig.3), one can compare
  (72) with results by Brndiar and Markos for  $n=5$ \cite{20} in
three dimensions (Fig.4,a). Due to the presence of  large
parameter $n(n-1)=20$, all numerical data lie in the critical
region  $x\alt 1$, where the dependence  $F_n(x)$ is practically
linear in accordance with  $\nu=1$ in the Vollhardt and W${\rm
{\ddot o}}$lfle theory.  Linearity of dependencies in Fig.4,a is
also evident, and their matching with the theoretical scaling
curve offers no problem (Fig.4,b).\,\footnote{\,Details of the
scaling procedure were discussed in Sec.3 of  \cite{5}.}

\begin{figure*}
\centerline{\includegraphics[width=6.0 in]{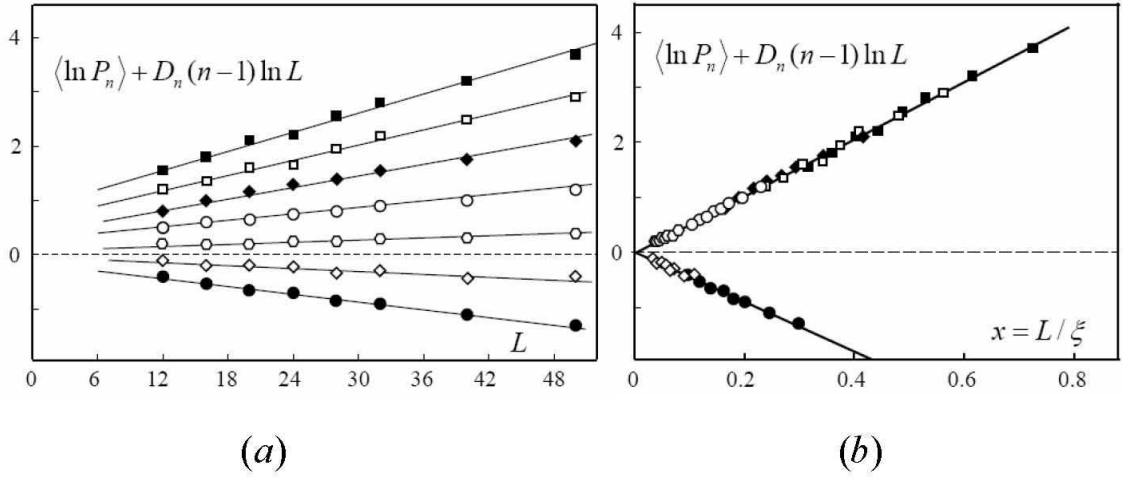}} \caption{
Numerical data by Brndiar and Markos for $n=5$ extracted from
Fig.2 in \cite{20}, and their comparison with the theoretical
scaling dependence. The empirical values  $D_2=1.28$ and
$D_5=0.96$ \cite{20} were used. A difference between
 $\langle \ln P_n\rangle$ and $\ln \langle P_n\rangle$ was
 neglected.  }
 \label{fig4}
 \end{figure*}
\begin{figure*}
\centerline{\includegraphics[width=6.0 in]{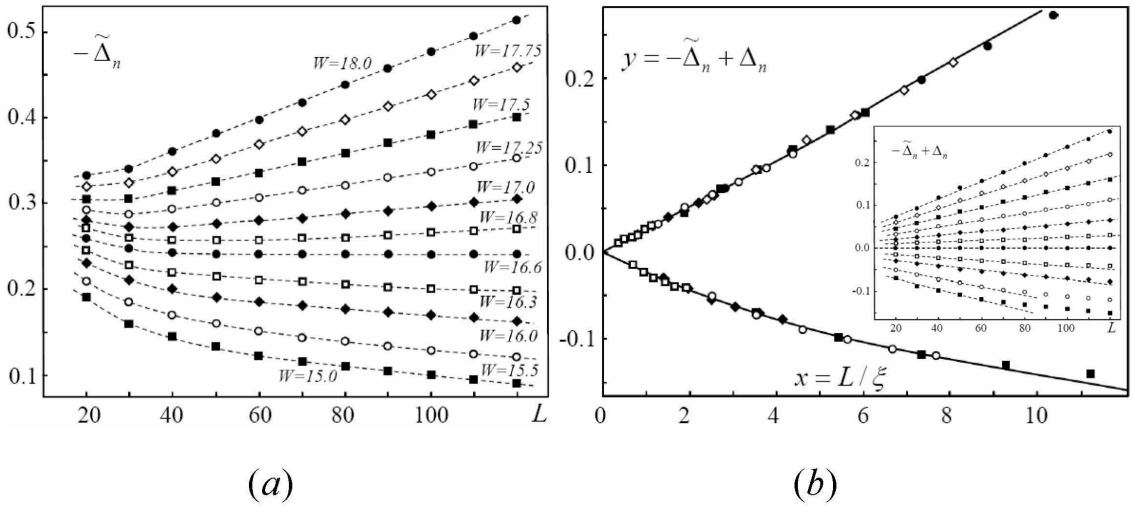}} \caption{
Numerical data by Rodriguez et al for $n=1/2$ extracted from
Fig.6,$c$  of \cite{21}, and their comparison with the
theoretical scaling dependence (72). Fig.5,$b$ contains points
corresponding to $L=20,\,40,\,60\ldots$.  }
\label{fig5}
\end{figure*}
The opposite situation takes place
for numerical data by Rodriguez et al \cite{21} for $n=1/2$
(Fig.5,a). In this paper the relation was accepted
$$
\langle P_n\rangle  \sim L^{-d(n-1)+\tilde\Delta_n}\,
 \eqno(74)
 $$
for the whole range of parameters, implying that
$\tilde \Delta_n= \Delta_n$ at the critical point; then
a comparison with (69) gives
$$
\tilde\Delta_n\,= \Delta_n + \frac{F_n(L/\xi)}{\ln(L/a)}
\,.
 \eqno(75)
 $$
In addition, roughening was made at the length scale $l$, which
should be used instead of $a$ in (75). If $\lambda=l/L$ is fixed,
then (75) contains only a dependence on $L/\xi$,
which is determined by the function $F_n(x)$.  Numerical data for
$n=1/2$, $\lambda=0.1$ \cite{21} are shown in Fig.5,a: the
stationary limit is reached for a value $W_c=16.6$, which gives
the estimate of the critical point. Accepting
$\tilde\Delta_n(W,L)-\tilde\Delta_n(W_c,L)$ as  deviation from
the critical behavior,  one can match all numerical data with the
theoretical scaling curve by a change of scale along the
horizontal axis (Fig.5,b). Due to existence of the small parameter
$n(n-1)/\ln(1/\lambda)=0.11$, the main body of data corresponds to
large values of  $x=L/\xi$, so the lower branch\,\footnote{\,Due
to the negative value of the factor $(n-1)$ the upper and lower
branch trade places, and to restore their natural disposition we
consider the quantity $-\tilde\Delta_n$. A definition of
$\tilde\Delta_n$ in \cite{21} was accepted with the opposite
sign, and Fig.5,a directly corresponds to Fig.6,c in \cite{21}.}
is determined by its logarithmic asymptotics, while the upper
branch remains in the linear regime due to a small value of
$\alpha$. It explains why dependencies for $W>W_c$ are
practically linear (see inset in Fig.5,b), while a tendency to
saturation is evident for $W<W_c$.  Small deviations in Fig.5,b
are probably related with inaccuracy of the interpolation form
(72). The evident linearity of dependencies at small $L$
corresponds to a value  $\nu=1$ of the Vollhardt and W${\rm
{\ddot o}}$lfle theory, while a statement of \cite{21} on the
result $\nu=1.590$ obtained with "unprecedented precision" looks
rather strange\,\footnote{\,The paper \cite{21} exploits the
treatment procedure developed in \cite{110}, which was already
criticized \cite{23}. It involves many-parameter nonlinear
fitting, which leads to the huge number of the $\chi^2$
minima and allows to obtain any desired value of $\nu$ in a
rather wide interval. A "desired" value $\nu=1.590$ was chosen
from the correspondence with preceding papers (occurrence of such
values was discussed in  \cite{1}), while its "unprecedented
precision" corresponds to fluctuations in the single  $\chi^2$
minimum and has no relation to actuality. Analogous arguments are
valid in respect to accuracy  of $\alpha_0$ (Sec.3) and
$D_2$ (Sec.5).}.

For frequencies $\omega\gg \Delta$, the following equations are
valid for $g$ and $z=L/\xi_{0D}$  \cite{3}
$$
g_L= \frac{p}{z^2} \,,\qquad
\pm c_d \left(L/\xi\right)^{d-2} =
\frac{p}{z^2}-c_d z^{d-2} \,,
\eqno(76)
$$
where $p=(-i\omega)/\Delta$. At the critical point one has
$\xi=\infty$ and  $z\sim p^{1/d}$, so $\xi_{0D}$ coincides with
the scale  $L_\omega$ introduced in (45). Therefore,  $m^{-1}\sim
L_\omega$ and the propagator  $\Pi({\bf r})$ falls exponentially
on the scale  $L_\omega\ll L$ providing statistical independence
of $|\psi_E({\bf r})|^2$ and $|\psi_{E+\omega}({\bf r})|^2$ for
$r\agt L_\omega$ and fulfilment of the normalization condition in
(44) apart to small deviations. At the critical point, Eqs.76 give
$g\sim \left(\omega/\Delta \right)^{(d-2)/d}$ in correspondence
with  \cite{13,14,15}; substituting to (44) and setting  ${\bf
r}={\bf r}'$, one has for small frequencies
$$
 \left\langle  |\psi_E({\bf r})|^2
|\psi_{E+\omega}({\bf r})|^2 \right\rangle \sim L^{-2d}
\left( \frac{L_\omega }{a} \right)^{\eta} \,
\propto \omega^{-\eta/d} \,.
\eqno(77)
$$
Numerical verification of such scaling was carried out
in papers  \cite{10,22} and was interpreted  as confirmation of
Chalker's hypothesis \cite{9} on a spatial dispersion of
$D(\omega,q)$. We see that this result can be obtained
without any reference to the $q$ dependence.

\begin{center}
{\bf 5. Convergence to the thermodynamic limit}
\end{center}

According to (61), all actual correlators are determined by
the diffusion propagator  $\Pi({\bf r})$ defined in (23), which
should be estimated for closed systems (Sec.2.4). The latter
contain ${\bf q}=0$ as an allowed value, and one can use the
periodical boundary conditions, accepting ${\bf q}=2\pi {\bf
s}/L$, where ${\bf s}$ is a vector with integer components
$s_i$.  For ${\bf r}\ne 0$, the sum over ${\bf q}$ is convergent
and no cut-off is necessary at large momenta.  Then one can
obtain\,\footnote{\,It follows from the $\alpha$-representation
and the Poisson summation formula (see Appendix in \cite{2}). }
$$
\Pi({\bf r})= \sum\limits_{\bf s} \Pi_0({\bf r}+{\bf s}L)   \,,
\eqno(78)
$$
where $\Pi_0({\bf r})$ is a continual version of (23)
$$
\Pi_0({\bf r}) = \int\, \frac{d^dq}{(2\pi)^d}
\frac{e^{i {\bf q}\cdot {\bf r}}} {{ q^2 + m^2}} =
$$
$$
=\,
\frac{2}{(4\pi)^{d/2}} \left( \frac{r}{2m} \right)^\mu
K_\mu(mr)\,, \qquad \mu=1-d/2 \,
\eqno(79)
$$
($K_\mu(x)$ is the Mac-Donald function) with the asymptotic
behavior for $d>2$:
$$
\Pi_0(r) =\left \{ \begin{array}{cc} \displaystyle{
 \frac{\Gamma(d/2-1)}{(4\pi)^{d/2}} m^{d-2}
\left( \frac{2}{mr} \right)^{d-2} } \,,  & mr\ll 1 \\
{                    }\\
\displaystyle{ \frac{(\pi/2)^{1/2}}{(2\pi)^{d/2}}
 \frac{m^{d-2}}{(mr)^{(d-1)/2}}\,e^{-mr} } \,, & mr\gg 1
\end{array} \right.\,,
  \,
\eqno(80)
$$
To provide a finite value for ${\bf r}=0$ we accept the spherical
cut-off $|{\bf q}|<\Lambda$, so
$$
\Pi_0(0) = \frac{K_d \Lambda^{d-2}}{d-2} \,
\eqno(81)
$$
and the growth at $r\to 0$ in (80) saturates for $r\alt
\Lambda^{-1}$. According to (78), $\Pi({\bf r})$ is
a sum of spherically symmetric functions originated in the
centers of cubical blocks with side $L$. This fact, along with
the cut-off $|{\bf q}|<\Lambda$, leads to  distortion of
dependencies (80) specific for infinite systems: it is manifested
 in anisotropy over directions of ${\bf r}$ and
in oscillations induced by the cut-off.  As a result, the
exponent in dependence $\Pi({\bf r})\propto |{\bf r}|^{-\alpha}$
is determined in a finite system with unavoidable restricted
accuracy.

According to \cite{2}, the relation $mL=z_0$ takes place at the
critical point, where $z_0$ is a root of the function $H(z)$
(see(65)). The parameter $z_0$ is not universal, but
depends on the details of cut-off, and hence on the specific
model ($z_0\approx 2$ for a spherical cut-off).  Fig.6,a
illustrates the results for $\Pi({\bf r})$ in the  $3D$ system of
size $L=20$ for $\Lambda=\pi$ and different $z_0$. One can see
that satisfactory power law dependencies $\Pi({\bf r})\sim |{\bf
r}|^{-\alpha}$ with $\alpha=0.27- 1.54$ take place for $z_0=1-4$,
\begin{figure}
\centerline{\includegraphics[width=3.0 in]{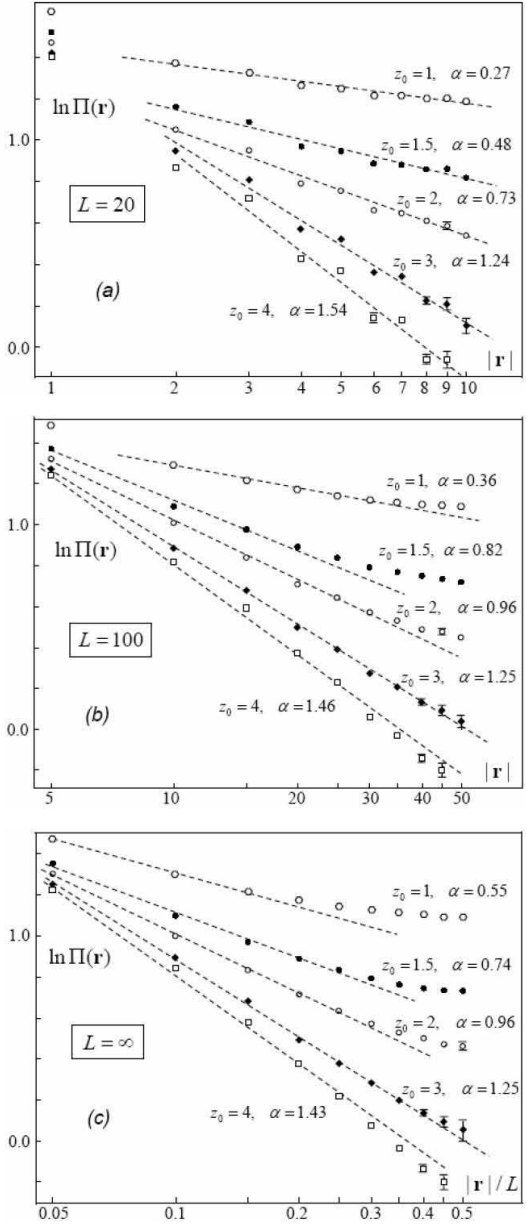}} \caption{
Behavior of the diffusion propagator  $\Pi({\bf r})$ with
different $z_0$ for  $L=20$ (a) $L=100$ (b) and
$L=\infty$ (c). The origin at the vertical axis is chosen
arbitrary.  }
\label{fig6}
\end{figure}
and their quality does not allow to distinguish a
theoretical value  $\alpha=1$.
If  $L$ is increased to 100 (Fig.6,b) the range of $\alpha$
becomes somewhat narrower for the same conditions, but remains
rather wide ($0.36-1.46$). Surprisingly,  the picture
does not change essentially even in the thermodynamic limit
$L\to\infty$ (Fig.6,c), if a value of $|{\bf r}|$ is a finite
fraction of $L$. Indeed, setting ${\bf r}={\bf y}L$, $mL=z_0$,
one has from (78) (the integration limits are shown for a
modulus of ${\bf q}$)
$$
\Pi({\bf r})= L^{2-d} \sum\limits_{\bf s}
\,\int\limits_0^{\Lambda L}\, \frac{d^dq}{(2\pi)^d}
\frac{e^{i{\bf q}\cdot {\bf(y+s)}}} {{ q^2 + z_0^2}}
\,,
\eqno(82)
$$
so that $\Lambda L\to\infty$ for large $L$, and the cut-off is
removed effectively; then $L$ enters only in the common factor
and does not affect the  ${\bf y}$ dependencies. The theoretical
value $\alpha=1$ should appear in the limit $|{\bf r}|/L\to 0$
independently of  $z_0$, but in fact this limit is unattainable
even for the maximal sizes  $L\sim 100$ reached in the present
time \cite{24,25}. One can see from Fig.6,c  that for $|{\bf r}|$
comparable with $L$ the exponent $\alpha$ is determined by the
$z_0$ value specific for a given model and has the scattering
$0.55-1.43$ for $z_0=1-4$. The satisfactory power law dependencies
are observed for  $|{\bf r}| > 0.05 L $, i.e. for
$|{\bf r}|> 5 $ at $L=100$, while the scales $|{\bf r}| \alt 5$
always fall out a scaling picture due to influence of
the cut-off. According to (61), $\Pi({\bf r})$ determines a
behavior of the $n$-point correlators, and it does not look
improbable if numerical estimations of fractal dimensions
may by tens of percents deviate from the true ones.  In
particular, it is hardly possible to make the statements of
principle, relying on deviations in the third digit  \cite{26}.

\begin{center}
\hspace{10mm} {\em T a b l e  1} \\
Estimations of $\eta$ and $D_2=d-\eta$ for the $3D$ systems of
different size\footnote{\,In respect of the last result
see Footnote 4 in \cite{18}}  $L$.

 \vspace{2mm}

\begin{tabular}{||c|c|c|c||}
\hline
 & &  &    \\
$L$     &  $\eta$  & $D_2$
&  Paper\\
  & & & \\
  \hline
  & & & \\
 10 & $1.4\pm 0.1$  & $1.6\pm 0.1$ & \cite{27}  \\

 16 & $1.3\pm 0.3$  & $1.7\pm 0.3$ & \cite{28}  \\

 20 & $1.67\pm 0.02$  & $1.33\pm 0.02$ & \cite{29}  \\

 40 & $1.3\pm 0.2$  & $1.7\pm 0.2$ & \cite{10}  \\

 40 & $1.5\pm 0.3$  & $1.5\pm 0.3$ & \cite{10}  \\

 47 & $1.32$  & $1.68$ & \cite{31}  \\

 48 & $1.48\pm 0.11$  & $1.52\pm 0.11$ & \cite{32}  \\

60 & $1.38\pm 0.18$ & $1.62\pm 0.18$  & \cite{30}  \\

80 & $1.70\pm 0.05$  & $1.30\pm 0.05$ & \cite{33}  \\

120 & $1.76\pm 0.03$  & $1.24\pm 0.03$ & \cite{21}  \\

240 & $1.76\pm 0.07$  & $1.24\pm 0.07$ & \cite{24}  \\

 & & & \\
 \hline
 \end{tabular}
 \end{center}
 \vspace{2mm}

According to the relation $\eta=2\alpha$ (see (58),
one can expect the scattering $\eta =1.1-2.8$. Table 1 gives
estimations of $\eta$ and $D_2=d-\eta$ obtained for $d=3$ by
different authors. One can see
their large scattering and a
systematic drift with increasing of $L$. The last estimate
$\eta=1.76\pm 0.07$ has only the $10\%$ deviation from the value
$\eta=2$ following from the first $\epsilon$-approximation, which
has a chance to be exact according to above arguments. The
observed deviations from the parabolical spectrum are also on the
level of  $10\%$ \cite{21,30,32}.  

\begin{center}
{\bf 6. On a spatial dispersion of the diffusion coefficient}
\end{center}

It is clear from the above discussion that all the picture related
with multifractality of wave functions can be obtained without
any reference to  the $q$-dependence of the diffusion coefficient
$D(\omega,q)$. At first glance, it indicates a
complete failure of Chalker's hypothesis \cite{9}. In fact, a
situation is more complicated due to ambiguity of the
$D(\omega,q)$ definition.

The arguments of \cite{9,10} are based on the relation
$$
{\cal K}(q) = \frac{\nu_F}{\pi} \,  \frac{D(\omega,q) q^2 }
{\omega^2  +\left[ D(\omega,q) q^2\right]^2}
\eqno(83)
$$
for the Fourier transform of correlator (19) and an
assumption on the similar behavior of correlators  ${\cal K}({\bf
r},{\bf r}')$ and $K({\bf r},{\bf r}')$ in the critical region
\cite{18}; then starting from $K({\bf r},0)\sim r^{-\eta}$
one has
${\cal K}(q)\sim K(q) \sim q^{-d+\eta}$ and
$D(\omega,q)\sim q^{d-2-\eta}$.  In
fact, the correct relation has a form (see below)
$$
{\cal K}(q) = \frac{1}{2\pi^2} \,{\rm Re} \left[
\frac{2 \pi \nu_F}{-i \omega + D(\omega, q) q^2} +
\phi_{reg} (q) \right]
\eqno(84)
$$
and reduces to (83) under assumption of the real diffusion
constant and irrelevance of the regular part  $\phi_{reg} (q)$.
The identical behavior of  ${\cal K}({\bf r},{\bf r}')$ and
$K({\bf r},{\bf r}')$ can be ensured in the limit of closed
systems (Sec.2.4) when $D(\omega, q)\propto (-i\omega)$ and the
pole term in (84) gives no contribution in the main order of
 $\omega$. In the general case,
  complex-valuedness
 of the diffusion coefficient does not allow to
 draw reliable conclusions relative $D(\omega, q)$ from the given
behavior of ${\cal K}(q)$.

According to \cite{17}, the use of the Kubo formulas
allows to relate the Fourier transform of (12) for ${\bf
r}_1={\bf r}_3$, ${\bf r}_2={\bf r}_4$ with the observable
diffusion coefficient
$$
\phi (q)  \, = \,
\frac{2 \pi \nu_F}{-i \omega + D(\omega, q) q^2} +
\phi_{reg} ( q) \,.
\eqno(85)
$$
Substitution of (85) into the expression for ${\cal K}({\bf
r},{\bf r}')$ analogous to (11) gives Eq.84 where the
regular part is somewhat different from (85) due to a
contribution of $\Phi^{RR}$.
  Decomposition into the pole and regular parts is
not unique and allows the  "gauge transformation"  \cite{17}
$$
{\tilde \phi}_{reg} (q) = \phi_{reg} (q) - 2 \pi
\nu_F C( q),
$$
$$
{\tilde D} (\omega,q) q^2 =
\frac{D (\omega, q) q^2 + i \omega C(q) [- i \omega +
D(\omega,  q) q^2]}{1 + C ( q) [-i \omega + D (\omega,
 q) q^2]},
\eqno(86)
$$
where  $C( q) =O(q^2) $ for small $q$. Another
representation for $\phi (q)$ follows from the spectral
properties of the quantum collision operator  \cite{17}:
if  $\lambda_s(q)$ are its eigenvalues, then
$$
\phi ( q) = \frac{A_0 (q)^2}{- \omega + \lambda_0 ( q)} +
\sum \limits_{s \neq 0} \frac{A_s ( q)^2}{- \omega +
\lambda_s (q)},
\eqno(87)
$$
where $A_0^2 ( q) = - 2 \pi i \nu_F + O( q^2)$, $ A_s (q)^2 =
O(q^2) $. The eigenvalue with  $s=0$ has a behavior $\lambda_0 (q)
\sim q^2$ for small $q$ and one can accept by definition
$$
\lambda_0 (q) = - iD (\omega,q) q^2  \,.
\eqno(88)
$$
Then (87) coincides with (85) where the regular part behaves
as $q^2$ for small  $q$ and can be excluded by the gauge
transformation. The gauge (88) will be referred as
"natural"; it was exploited in \cite{17} and  found to
be free of an essential spatial dispersion. Another
distinguished gauge is defined by the condition
$\phi_{reg} (q)\equiv 0$;  it is actual in the
localized phase, where $D (\omega,q) = (-i\omega) d(q)$
and the following relation
$$
\frac{1}{1 + d(q)q^2} = {\cal A} ({\bf q})
= \int d {\bf r} e^{-i {\bf q} \cdot {\bf r}}
{\cal A} ({\bf r})
$$
$$ {\cal A} ({\bf r}) =\frac{1}{\nu_F}
\left\langle \sum \limits_s | \psi_s ({\bf r})|^2 |
\psi_s (0)|^2 \delta (E - \epsilon_s) \right\rangle\,
\eqno(89)
$$
can be obtained from the Berezinskii-Gor'kov criterion \cite{17}.
The term with $s=s'$ in correlator (19) gives contribution
$\delta(\omega)$ in the localized phase, transforming
to the singularity $1/\omega$ in the quantity $\phi(q)$, which
can be identified with the diffusion pole in (85). It is
of principle importance to gather all contributions  $\sim 1/\omega$
which may be contained in $\phi_{reg}(q)$ and to include them
in the pole term. They are certainly present in $\phi_{reg}(q)$
for gauge (88), since there are terms with
$\lambda_s(q)\sim \omega$ in the sum of (87) \cite{17}. Such
contributions are surely included, if the gauge with
$\phi_{reg}(q)\equiv 0$ is chosen, and just this gauge is
implied in (89). Comparison of (89) with (4) shows that
${\cal A}(r)\sim r^{-\eta}$ for  $r\alt \xi$ and
$d(q)\sim q^{d-2-\eta}$ for $q\agt \xi^{-1}$  in correspondence
with the Chalker hypothesis\,\footnote{\,The results
for $D(\omega,q)$ in the
localized phase and the critical region are matched at  $\xi\sim
L_{\omega}$; thus at the critical point one has
$D(\omega,q)\sim \omega L^2_{\omega} \sim \omega^{(d-2)/d}$
for  $q L_{\omega} \alt 1$  and  $D (\omega,q)\sim
\omega L^2_{\omega} (q L_{\omega})^{d-2-\eta} \sim
\omega^{\eta/d} q^{d-2-\eta}$  for  $q L_{\omega} \agt 1$.},
while $d(q)=const=\xi^2$ for
$q\alt \xi^{-1}$ and  ${\cal A}(r)\sim \exp(-r/\xi)$ for
$r\agt \xi$.  If the gauge with $d(q)=const$ is used,
then the contribution $q^{-d+\eta}/\omega$ is contained in
$\phi_{reg}(q)$.

The "natural" gauge  (88) was used in the analysis of \cite{17}
and just that very definition of $D(\omega,q)$ is implied in the
vertex  $U^{RA}$. If the pole approximation is used in Eq.12 of
\cite{17}, then one can set  ${\bf k}'=-{\bf k}$ in the function
$F({\bf k}, {\bf k}', {\bf q})$ and obtain $F({\bf k},-{\bf k},
{\bf q})=2U_0\gamma$ from Eq.65 of this paper, if  ${\rm Im}
\Sigma^R_{\bf k}=-\gamma$ is accepted to be ${\bf k}$-independent
and $U_0$ is defined by the relation $\gamma=\pi U_0 \nu_F$. Then
the pole term of the vertex $U^{RA}$ can be obtained from the
cooperon contribution (14) by substitution of $D(\omega,q)$ for
$D_0$ and neglecting the $q$ dependence. The precisely such form
of the vertex was used in the above considerations, and its
validity is confirmed by successful reproducing of  multifractal
properties. It  gives an essential support to the conclusions
of \cite{17}.

\begin{figure*}
\centerline{\includegraphics[width=5.0 in]{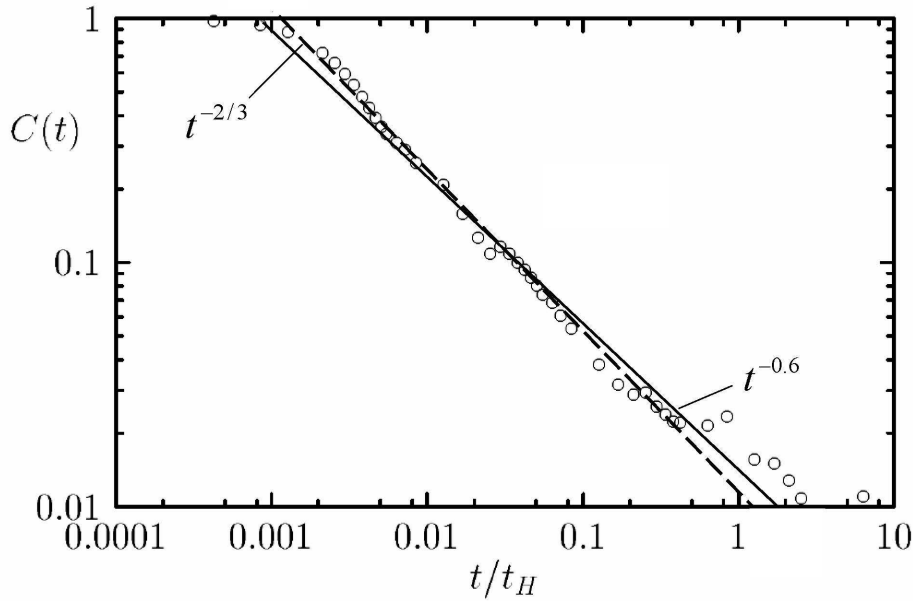}} \caption{
The raw data by Brandes et al \cite{10} on spreading of the
wave packet; the autocorrelation function $C(t)\sim t^{-(d-\eta')/d}$
describes the change of the amplitude in the packet center
as a function of time. Dependence $t^{-2/3}$ corresponds
to the absence of a spatial dispersion of the diffusion
coefficient, while dependence $t^{-0.6}$ is the result
indicated in  \cite{10}.  }
\label{fig7} \end{figure*}

The paper \cite{17} contains inaccuracy consisting in a wrong
interpretation of the "regularity" of $\phi_{reg}(q)$. It was
suggested that the Anderson transition is completely determined by
diffusion poles, while the function $\phi_{reg}(q)$ does not feel
the transition and contains no information on the correlation
length $\xi$. Then the quantity $C(q)$ in (86) relating two such
regular functions is also $\xi$-independent, and
absence of the anomalous dispersion (related with the scale
$\xi$) in one gauge leads to its absence in another gauge. In
fact, as we see, information on $\xi$ is unavoidably present
either in $D(\omega,q)$, or in $\phi_{reg}(q)$.

The latter leads to the disappointing conclusion, that
nothing can be said on the gauge corresponding
to the observable diffusion constant, so the exponent  $\eta'$
in (9) is in general different from
$\eta$. The most detailed verification of the relation
$\eta'=\eta$ was carried out in the paper \cite{10}. In fact, a
value  $\eta'=1.20\pm 0.15$ was found for the exponent $\eta'$
and two estimates ($\eta=1.3\pm0.2$ and $\eta=1.5\pm 0.3$)
were obtained for the exponent $\eta$: according to the authors,
it was sufficient to establish the equality  $\eta'=\eta$.
Lately, the estimate for $\eta$ has drifted to 1.76 (see
Table 1), while no fresh data for  $\eta'$
is known to us. One can see from Fig.7 that the raw
data of \cite{10} for the autocorrelation dependence
$t^{-(d-\eta')/d}$ are perfectly compatible with a value
$\eta'=1$ corresponding to absence of the spatial dispersion. The
physical experiment on spreading of the wave packet \cite{34} is
also well-described by self-consistent theory of localization.

\begin{center}
{\bf 7. On deficiency of  $\sigma$-models}
\end{center}

In Sec.3 we have established deficiency of the
$\sigma$-model approach beyond the first
$\epsilon$-approximation.  This situation is not
unexpected: derivation of $\sigma$-models is justified only
for small  $\epsilon$, and the question on their exact
correspondence with the initial disordered systems always
remained open. In particular, strong doubts arouse in relation with the
upper critical dimension \cite{5}. Here we present arguments, why
 deficiency of $\sigma$-models for the orthogonal
ensemble arises just on the four-loop level.

Following arguments of Sec.3, we can assume,
that the two-cooperon form of correlator $K({\bf r},{\bf r}')$
is exact. Then  absence of the spatial dispersion of
$D(\omega,q)$ corresponds to the exact equality  $\eta=2\epsilon$,
while violation of this equality corresponds to  appearance
of the spatial dispersion.

Wegner's result (3) is obtained in the "minimal" $\sigma$-model,
where the lowest (second) powers of gradients are retained: it
corresponds to neglecting a spatial dispersion of $D(\omega,q)$.
In the first three orders in $\epsilon$, the equality
$\eta=2\epsilon$ takes place and the approximation is
self-consistent. This equality is violated on
the four-loop level and signals on the lack of self-consistency.
The terms with higher gradients should be taken into account,
which leads to instability of the renormalization group due to
the "gradient catastrophe" \cite{35}.  To remove instability,
additional counter-terms should be added; it leads to essential
modification of the  $\sigma$-model Lagrangian and makes unclear
a fate of the four-loop contribution.  It should be stressed that
according to the analysis of the paper \cite{17} the spatial
dispersion is determined by the atomic scale. It is inessential
in the practical sense but its existence is a matter of principle
due to the infinite-component nature of the order parameter.

There is another evidence of the $\sigma$-model deficiency. If the
Vollhardt and W$\ddot o$lfle  theory is exact, then the formalism
of dimensional reqularization is initially incompatible with the
physical essence of the problem \cite{2}. The enforced
application of this formalism should lead to unsolvable problems,
and the "high-gradient catastrophe" is a possible manifestation of
them. This catastrophe is probably eliminated in other
regularization schemes  (see  discussion of the paper \cite{37}
in  \cite{36}), but a change of the scheme surely modifies
many-loop contributions.

We should stress that discrepancy between $\sigma$-models
and the self-consistent theory of localization arises just on
the four-loop level. There is a chance for elimination of
this discrepancy in the result of indicated modifications of
$\sigma$-models.

\begin{center}
{\bf 8. Conclusion}
\end{center}

We have shown above that multifractal properties of wave functions
can be obtained from self-consistent theory of localization by
Vollhardt and W${\rm {\ddot o}}$lfle, in spite of the opposite
statements widespread in literature. The diagrammatic
interpretation of results allows  to derive all scaling relations
used in numerical experiments. Comparison with the latter confirms
the tendency revealed in preceding papers \cite{1,2,3,4,5}: the
raw numerical data are perfectly compatible with the Vollhardt and
W${\rm {\ddot o}}$lfle theory, while the opposite statements of
the original papers are related with ambiguity of interpretation
and existence of small parameters of the Ginzburg number type.

Analysis of the first $\epsilon$-approximation of the $2+\epsilon$
theory reveals existence of two qualitative properties:  (i)
irrelevance of the correlation length $\xi$ in the metallic phase
as a characteristic length scale, and (ii) realization of the
maximally symmetric form (56) for the $n$-point correlator of wave
functions. Due to a qualitative character of these properties they
have a chance to be exact; then the multifractal spectrum is
strictly parabolic and determined by the one-loop Wegner result. A
surprising accuracy of this result in application to the  $d=3$
and $d=4$ cases was repeatedly reported in literature, while the
observed small deviations can be explained by the extremely slow
convergence to the thermodynamic limit discovered in Sec.5. The
four-loop contribution to the anomalous dimensions is surely
deficient and may disappear in result of a modification of the
$\sigma$-model Lagrangian, which is necessary for  taking into
account the spatial dispersion of $D(\omega,q)$ and elimination
the gradient catastrophe. Simultaneously, it may lead to
elimination of other discrepancies between $\sigma$-models and the
self-consistent theory, which are present on the four-loop level.
As noted in  Sec.3,  validity of the self-consistent theory is
directly connected with the property  (i).

As for the relation of multifractality with a spatial dispersion
of the diffusion coefficient  $D(\omega,q)$, this question is
resolved in the compromise manner. A definition of $D(\omega,q)$
is ambiguous and allows the "gauge transformation". A spatial
dispersion is absent in the "natural" gauge (88),  while the
Chalker hypothesis \cite{9} takes place in the gauge with
$\phi_{reg}(q)\equiv 0$ . The raw data of paper  \cite{10} on spreading
of the wave packet and the physical experiment \cite{34}
indicate the absence of a spatial dispersion for the
observable diffusion constant.


\begin{thebibliography}{xx}


\bibitem{1}  I. M. Suslov,  Zh. Eksp. Teor. Fiz. {\bf 141},
122 (2012)   [J. Exp. Theor. Phys. {\bf 114}, 107  (2012)].

\bibitem{2}  I. M. Suslov, Zh. Eksp. Teor. Fiz. {\bf 142},
1020 (2012)    [J. Exp. Theor. Phys. {\bf 115}, 897  (2012)].

\bibitem{3}  I. M. Suslov, Zh. Eksp. Teor. Fiz. {\bf 142}, 1230
(2012) [J. Exp. Theor. Phys. {\bf 115}, 1079  (2012)].

\bibitem{4} I. M. Suslov, Zh. Eksp. Teor. Fiz. {\bf 145}, 1031 (2014)
[J. Exp. Theor. Phys. {\bf 118}, 909  (2014)].

\bibitem{5} I. M. Suslov, Zh. Eksp. Teor. Fiz. {\bf 146}, 1272 (2014)
[J. Exp. Theor. Phys. {\bf 119}, 1115  (2014)].

\bibitem{6}  P. Markos, acta physica slovaca 56, 561 (2006);
cond-mat/0609580.

\bibitem{7} D. Vollhardt, P. W$\ddot o$lfle, Phys. Rev. B {\bf
 22}, 4666 (1980); Phys. Rev. Lett. {\bf 48}, 699 (1982).

\bibitem{100} H. Kunz, R. Souillard, J. de Phys. Lett. {\bf 44},
L506 (1983).

\bibitem{17} I. M. Suslov, Zh. Eksp. Teor. Fiz. {\bf 108},
1686 (1995)    [J. Exp. Theor. Phys. {\bf 81}, 925 (1995)].

\bibitem{8}  F. Wegner, Nucl. Phys. B 316, 663 (1989).

\bibitem{9} J.~T.~Chalker, Physica~A {\bf 167}, 253  (1990).

\bibitem{10} T. Brandes, B. Huckestein, L. Schweitzer,
Ann. Phys. {\bf 5}, 633 (1996).

\bibitem{11} E. Abrahams, P. W. Anderson, D. C. Licciardello, and
T. V.  Ramakrishnan, Phys. Rev. Lett. {\bf 42}, 673 (1979).

\bibitem{12}  I. M. Suslov, cond-mat/0612654

\bibitem{12a}  A. Kawabata,  cond-mat/0104289

\bibitem{101}  A. M. Garcia-Garcia,  Phys. Rev. Lett.
 {\bf 100}, 076404 (2008).

\bibitem{102} N. N. Bogoliubov and D. V. Shirkov, Introduction to the
Theory of Quantized Fields (Nauka, Moscow, 1976;
Wiley, New York, 1980).

\bibitem{13}  F. Wegner, Z. Phys. B 25, 327 (1976).

\bibitem{14} B. Shapiro, E. Abrahams, Phys. Rev. B {\bf 24},
4889 (1981).

\bibitem{15} S. Hikami, Phys. Rev. B {\bf 24}, 2671 (1981).

\bibitem{16} P. Lambrianides, H. B. Shore,  Phys. Rev.
  B {\bf 50}, 7268 (1994).

\bibitem{103} M. V. Sadovskii, Diagrammatics, World Scientific,
Singapore, 2006.

\bibitem{18}  M. V. Feigelman, L. B. Ioffe, V. E. Kravtsov,
E. Cuevas,  Annals of Physics (NY) {\bf 325}, 1368 (2010).

\bibitem{19} F. Evers, A. D. Mirlin, Rev. Mod. Phys. {\bf 80},
1355 (2008).

\bibitem{105} I. M. Suslov, Usp. Fiz. Nauk {\bf 168},  503  (1998)
[Physics -- Uspekhi {\bf 41}, 441 (1998)].

\bibitem{33} A . M. Mildenberger, F. Evers, A. D. Mirlin,
Phys.  Rev.  B {\bf 66}, 033109 (2002).

\bibitem{21} A. Rodriguez, L. J. Vasquez, K. Slevin, R. A. Romer,
Phys. Rev. B {\bf 84}, 134209 (2011).

\bibitem{30} H. Grussbach, M. Schreiber, Phys.  Rev.  B
{\bf 51}, 663 (1995).

\bibitem{32} T. Terao,  Phys.  Rev. B  {\bf 56}, 975 (1997).

\bibitem{26} F. Evers,  A. M. Mildenberger,  A. D. Mirlin,
Phys.  Rev.  Lett. {\bf 101}, 116803 (2008).


\bibitem{107} M. Zirnbauer, hep-th/9905054.

\bibitem{108}  M. J. Bhasen, et al, Nucl. Phys. B {\bf 580},
688 (2000).

\bibitem{109} A. M. Tsvelik, Phys. Rev. B {\bf 75},
184201 (2007).

\bibitem{150} I. M. Suslov, arXiv: 1412.5339.

\bibitem{20} J. Brndiar, P. Markos, Phys. Rev. B {\bf 74},
153103 (2006).

\bibitem{110} K. Slevin, T. Ohtsuki, Phys. Rev. Lett. {\bf 82},
382 (1999).  

\bibitem{23} I.~M.~Suslov, cond-mat/0105325, 0106357.

\bibitem{22} E. Cuevas, V. E. Kravtsov,  Phys. Rev. B {\bf 76},
235119 (2007).

\bibitem{27} M. Schreiber, Physica A {\bf 167}, 188 (1990).

\bibitem{28} C. M. Soukoulis, E. N. Economou, Phys. Rev. Lett.
 {\bf 52}, 565 (1984).

\bibitem{29} S. N. Evangelou, Physica A {\bf 167},
199 (1990).

\bibitem{31} M. Schreiber, H. Grussbach,  Phys.  Rev. Lett.
{\bf 67}, 607 (1991).

\bibitem{24} A. Rodriguez, L. J. Vasquez,  R. A. Romer,
Phys. Rev. Lett. {\bf 102}, 106406-4 (2009).

\bibitem{25} A. Rodriguez, L. J. Vasquez,  R. A. Romer,
Phys. Rev. B {\bf 78}, 195107 (2008).

\bibitem{34} G. Lemarie, H. Lignier, D. Delande, et al,
arXiv:1005.1540

\bibitem{35}  V. E. Kravtsov, I. V. Lerner, V. I. Yudson,
Zh. Eksp. Teor. Fiz. {\bf 94}, 255 (1988) [Sov. Phys. JETP
{\bf 67}, 1441 (1988)].

\bibitem{37} P. K. Mitter, H. R. Ramadas, Commun. Math. Phys.
{\bf 122}, 575 (1989).

\bibitem{36}  F. Wegner, Z. Phys. B {\bf 78}, 33 (1990).



\end{thebibliography}
\end{document}